\documentclass[11pt,a4paper]{article} 
\pdfoutput=1
\usepackage[whole]{bxcjkjatype} 
\usepackage{amsmath}
\usepackage{amssymb}
\usepackage{amsthm}

\usepackage{setspace}
\usepackage{empheq}
\usepackage{bm}
\usepackage{color}
\usepackage{cite} 
\usepackage{braket}
\usepackage{float}

\theoremstyle{definition}

\usepackage{hyperref}
\usepackage{graphicx}

\catcode `@=11 \@addtoreset{equation}{section} \catcode `@=12

\begin{document}
\setlength{\oddsidemargin}{0cm}

\begin{titlepage}
\begin{flushright} 
	KEK-TH-2846
\end{flushright} 

    \begin{center}
        {\LARGE	
Underdetermination in the Quantum-Mechanical Bootstrap \\
 with Mixed Functional Operators
		}
    \end{center}
    \vspace{1.2cm}
    \baselineskip 18pt 
    \renewcommand{\thefootnote}{\fnsymbol{footnote}}

    \begin{center}
        Takeshi {\sc Morita}$^{a, b}$\footnote{%
            Contact author: morita.takeshi(at)shizuoka.ac.jp
        }, 
        Worapat {\sc Piensuk}$^{c, d}$\footnote{%
        	Contact author: piensukw(at)ibs.re.kr        
},
        and Pushkar {\sc Soni}$^e$\footnote{
            Contact author: pushkars21(at)iitk.ac.in
        }
        
        \renewcommand{\thefootnote}{\arabic{footnote}}
        \setcounter{footnote}{0}
        
        \vspace{0.4cm}
        
        {\it
			a. Department of Physics,
            Shizuoka University \\
            836 Ohya, Suruga-ku, Shizuoka 422-8529, Japan
        \vspace{0.2cm}
        \\    
		b. Graduate School of Science and Technology, Shizuoka University\\
		836 Ohya, Suruga-ku, Shizuoka 422-8529, Japan
            \vspace{0.2cm}
            \\
        c. The Graduate University for Advanced Studies, SOKENDAI\\
        1-1 Oho, Tsukuba, Ibaraki 305-0801, Japan
        \vspace{0.2cm}
        \\
        d. KEK Theory Center, Institute of Particle and Nuclear Studies,\\
        High Energy Accelerator Research Organization, \\
        1-1 Oho, Tsukuba, Ibaraki 305-0801, Japan
        \vspace{0.2cm}
        \\
        e. Indian Institute of Technology Kanpur, Kanpur 208016, India
        }

    \end{center}

    
    \vspace{1.5cm}

\begin{abstract}

The bootstrap method for quantum mechanics is a powerful tool for computing the energy eigenvalues of a Hamiltonian. However, we point out that a conventional implementation employing the standard bootstrap matrix may be insufficient when the potential contains different types of functions, such as polynomial and exponential terms. In such cases, the standard constraints may not fully capture the relations between the corresponding operator sectors, and the method may fail to reproduce the correct spectrum. A similar problem may arise when evaluating expectation values of operators of different types. This issue can occur in a wide range of systems, including statistical models and matrix models. We propose three possible resolutions to this problem.

\end{abstract}

    
\end{titlepage}

\tableofcontents

\section{Introduction}
\label{sec-Intro}

In recent years, numerical analyses based on the bootstrap method have attracted significant attention \cite{Anderson:2016rcw, Lin:2020mme, Han:2020bkb}. This approach determines the allowed regions of physical observables that satisfy positivity of probabilities as well as system-specific constraints, such as the Schwinger-Dyson equations. Related positivity constraints were used earlier in the eigenvalue moment method to bound energy eigenvalues \cite{PhysRevLett.55.931, PhysRevA.36.4411, HANDY1987308, PhysRevA.37.4557, PhysRevLett.60.253, PhysRevA.52.3468}. The bootstrap method has been successfully applied to a wide range of systems, including matrix models \cite{Lin:2020mme, Han:2020bkb, Kazakov:2021lel, Lin:2023owt, Li:2024ggr, Khalkhali:2020jzr, Cho:2024kxn, Lin:2024vvg, Lin:2025srf, Lin:2025iir, Kovacik:2025qgj, Laliberte:2025xvk, Li:2025tub, Cho:2025vws, Berenstein:2026wky}, quantum mechanics \cite{Han:2020bkb, Berenstein:2021dyf, Bhattacharya:2021btd, Aikawa:2021eai, Berenstein:2021loy, Tchoumakov:2021mnh, Aikawa:2021qbl, Du:2021hfw, Bai:2022yfv, Nakayama:2022ahr, Li:2022prn, Khan:2022uyz, Hu:2022keu, Berenstein:2022ygg, Morita:2022zuy, Blacker:2022szo, Berenstein:2022unr, Berenstein:2023ppj, Sword:2024gvv, Li:2024rod, Lawrence:2024mnj, Berenstein:2025itw, Aikawa:2025dvt, Huang:2025sua, Lawrence:2025wyl, Thong:2026zvt}, many-body systems \cite{Bhattacharya:2021btd, Morita:2022zuy, Cho:2022lcj, Nancarrow:2022wdr, Lawrence:2022vsb, Cho:2023ulr, Berenstein:2024ebf, Gao:2024etm, Cho:2024owx, Cho:2025dgc, Chowdhury:2025dlx, Cho:2026lfu}, and lattice field theories \cite{Anderson:2016rcw, Lawrence:2021msm, Kazakov:2022xuh, Kazakov:2024ool, Li:2024wrd, Guo:2025fii}.

One of the key advantages of this method is its ability to treat the large-$N$ and infinite-volume limits of many-body systems \cite{Anderson:2016rcw, Lin:2020mme, Han:2020bkb}. Furthermore, by employing the Hamiltonian formalism, it is expected to circumvent the sign problem that often plagues Monte Carlo simulations \cite{Aikawa:2021eai}. In one-dimensional quantum mechanics, the method yields energy eigenvalues with high precision, and it can even reproduce exact spectra for solvable systems \cite{Aikawa:2021qbl, Sword:2024gvv, Aikawa:2025dvt}.

Despite these successes, the theoretical foundations of the bootstrap method have yet to be fully elucidated. In this paper, we point out a theoretical limitation of a conventional implementation employing the standard bootstrap matrix. Specifically, we demonstrate that this implementation may be insufficient when multiple types of functions, such as combinations of polynomial and exponential terms, are involved. We show that this failure originates from an underdetermination of the relations between the corresponding operator sectors.

As a concrete example, we consider one-dimensional quantum mechanics with a Hamiltonian whose potential is a mixture of polynomial and exponential terms. We apply this conventional implementation to the system and show that it fails to reproduce the energy spectrum. We then provide possible prescriptions for resolving this issue. Although these prescriptions work for the specific example considered here, they may not provide a universal solution. 

Crucially, this problem is not restricted to one-dimensional quantum mechanics; it can also arise in other systems, such as statistical models. This presents a challenge in applying the bootstrap approach to such systems.\\

The organization of this paper is as follows. In Section~\ref{sec-ambiguity}, we explain the underdetermination between different operator sectors in a conventional implementation of the bootstrap method, using a Hamiltonian with a mixed potential as a concrete example. In Section~\ref{sec-ambiguity-resolution}, we propose possible resolutions to this problem. Finally, in Section~\ref{sec-discussion}, we summarize our results and discuss future directions. In Appendix~\ref{appen-sec:recursion-derivation}, we derive the relations among the expectation values of the operators. In Appendix~\ref{appen-sec:bootstrap-method-detail}, we describe the numerical methods used in this paper. 
In Appendix~\ref{appen-sign}, we discuss additional resolutions to the underdetermination associated with the operator $e^{\hat{x}}$.

We use units with $\hbar=1$ throughout this paper.

\section{Underdetermination in a Conventional Implementation of the Bootstrap Method}
\label{sec-ambiguity}

\subsection{Failure of the Conventional Bootstrap Matrix for a Mixed Potential}
\label{sec-ambiguity-mixed}

To demonstrate the underdetermination in a conventional implementation of the bootstrap method, we consider the following two systems. One is the anharmonic oscillator (AHO), whose Hamiltonian is given by
\begin{align}
    H_{\rm AHO}=\frac{1}{2}\hat{p}^2+ \frac{1}{2}\hat{x}^2 + \frac{1}{4} \hat{x}^4 .
    \label{eq-H-anharmonic}
\end{align}
The other is the system defined by
\begin{align}
    H_{\rm mixed}=\frac{1}{2}\hat{p}^2 + (\hat{x}-L)^2 +\cosh(\hat{x}) ,
    \label{eq-H-coshx}
\end{align}
where $L$ is a real constant.
The potential of this Hamiltonian contains both polynomial and exponential terms, and thus we refer to it as a mixed potential.

The bootstrap method has been applied to the anharmonic oscillator \eqref{eq-H-anharmonic} in the literature \cite{Han:2020bkb}, and it successfully reproduces the energy spectrum of this system.
In contrast, we will see that the same implementation fails to reproduce the energy spectrum of the mixed potential \eqref{eq-H-coshx}.
Later, we will compare these two systems and explain why this implementation works for the anharmonic oscillator but not for the mixed potential.
\\

\begin{figure}[htbp]
    \centering
    \includegraphics[width=0.8\textwidth]{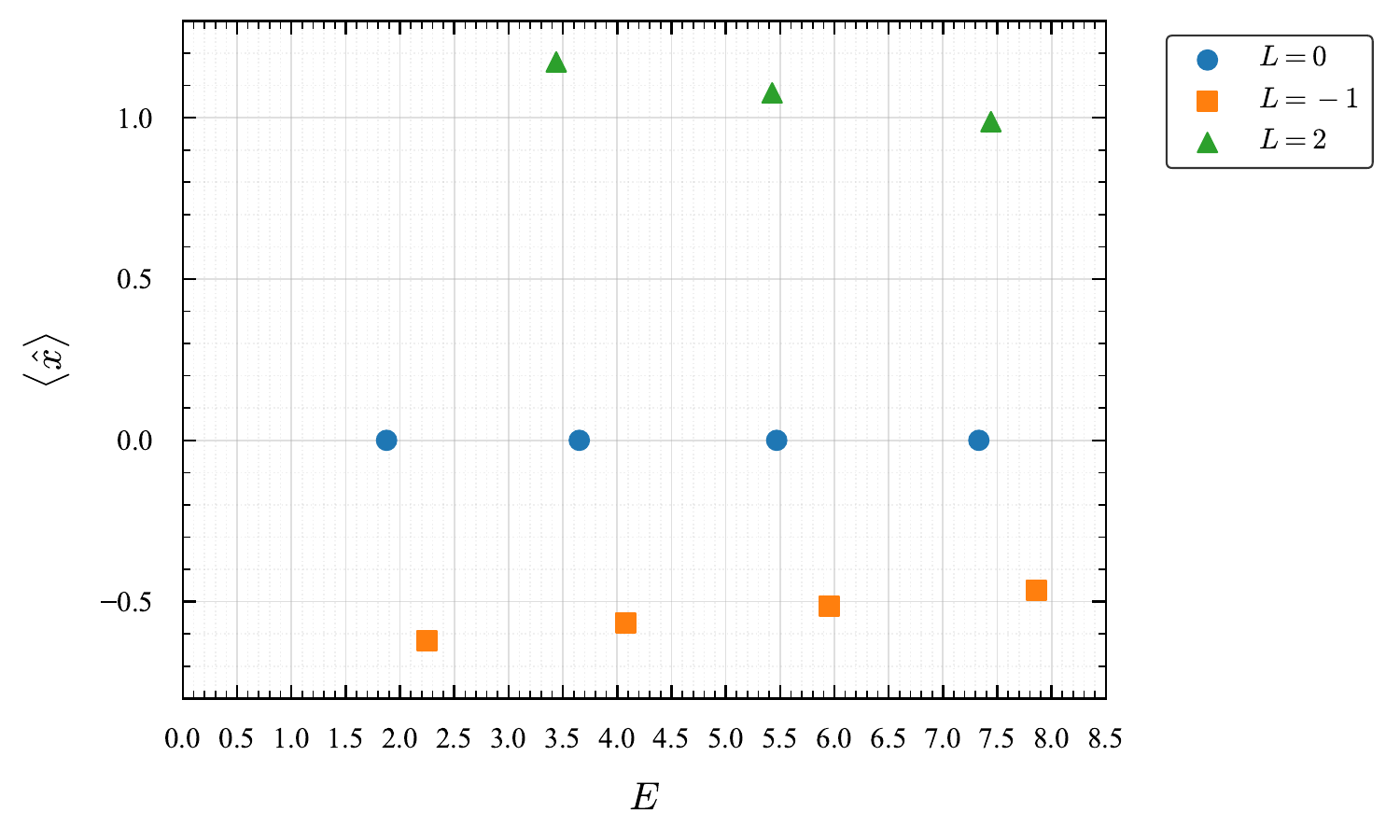}
    \caption{
    The spectrum of the Hamiltonian with the mixed potential \eqref{eq-H-coshx} for $L = 0, -1,$ and $2$.
    We plot the energy eigenvalues $E$ against the expectation values $\langle \hat{x} \rangle$.
    The spectrum clearly depends on the value of $L$.
    We use the \texttt{NDEigensystem} package in \textit{Mathematica} to solve the Schr\"odinger equation numerically, and we later compare this spectrum with the results obtained using the bootstrap method.
    }
    \label{fig-spectrum}
\end{figure}

Before analyzing these systems using the bootstrap method, let us first examine the spectrum of the mixed potential. We numerically solve the Schr\"odinger equation for the Hamiltonian \eqref{eq-H-coshx} and obtain the energy eigenvalues $E$ and the expectation values $\braket{\hat{x}}$ at the corresponding eigenstates. We then plot them in the $(E,\Braket{\hat{x}})$-plane, as shown in Fig.~\ref{fig-spectrum}.
We consider several values of $L$ ($L=0,-1$ and $2$), and we use the \texttt{NDEigensystem} package in \textit{Mathematica} to solve the Schr\"odinger equation numerically.
At $L=0$, the potential has a parity symmetry $\hat{x} \to -\hat{x}$, and therefore $\braket{\hat{x}}=0$ for all energy eigenstates.
Moreover, the spectrum depends on the value of $L$. We will see that the implementation considered here fails to reproduce this $L$-dependence and consequently cannot reproduce the spectrum of this system.
\\

Now, we study the bootstrap method for these two systems. Suppose that $P$ is any positive semidefinite operator, so that $\braket{P}\geq 0$ for any state. Then, the following positivity condition holds:
\begin{align} 
  \braket{\tilde{O}^\dagger P \tilde{O}} \ge 0, \quad \tilde{O}= \sum_{i=1}^K c_i O_i .
  \label{eq-positivity}
 \end{align}
 Here, $\braket{\cdots}$ denotes the expectation value in an energy eigenstate, $c_i$ are arbitrary complex coefficients, and $\{ O_i \}$ $(i=1,2,\ldots,K)$ is a set of operators arranged in some order.
 Their ordering does not affect the result.
Since $c_i$ are arbitrary, this positivity condition is equivalent to requiring that the $K \times K$ Hermitian matrix ${\mathcal M}$ defined by
\begin{align}
	{\mathcal M}_{ ij}:=
		\left\langle O_i^\dagger P O_j \right\rangle , 
	\label{eq-bootstrap-matrix}
\end{align}
be positive semidefinite, ${\mathcal M} \succeq 0$.
This matrix ${\mathcal M}$ is called a bootstrap matrix.

To proceed, we need to choose the positive semidefinite operator $P$ and the operator set $\{ O_i \}$.
A simple choice is to take $P=1$ and construct $\{O_i\}$ from the operators appearing in the Hamiltonian. Thus, we take
\begin{align} 
& \{\hat{x}^m  \hat{p}^n  \}, \quad m=0,1, \cdots, K_x,~ n=0,1, \cdots, K_p,  &\text{(anharmonic oscillator)},  
 \label{eq-operators-XP} \\
 &   \{\hat{x}^l  e^{m\hat{x}} \hat{p}^n  \}, \quad l=0,1, \cdots, K_x,~m=-K_{\rm e}, \cdots, K_{\rm e},~ n=0,1, \cdots, K_p,  &\text{(mixed potential)}.  
 \label{eq-operators-XEP}
 \end{align}
Here $K_x$, $K_{\rm e}$ and $K_p$ are non-negative integers, and $K$ is given by $K= (K_x+1)(K_p+1)$ for the anharmonic oscillator and $K= (K_x+1)(2K_{\rm e}+1)(K_p+1)$ for the mixed potential. In this paper, we refer to the matrix constructed with $P=1$ and the operator sets \eqref{eq-operators-XP} and \eqref{eq-operators-XEP} as the conventional bootstrap matrix, and we refer to the corresponding implementation as the conventional bootstrap method\footnote{In the original work in Ref.~\cite{Han:2020bkb}, the operator set $\{ O_i \}= \{ x^i \}$~\cite{PhysRevA.36.4411, HANDY1987308} is taken instead of \eqref{eq-operators-XP} for the anharmonic oscillator. We take the operator sets \eqref{eq-operators-XP} and \eqref{eq-operators-XEP}, which involve the momentum operator $\hat{p}$, because they are more general and may provide stronger constraints.}.

For example, in the case of the anharmonic oscillator,
${\mathcal M}$ is given by
\begin{align}
    \mathcal{M}=\begin{pmatrix}
1 & \langle \hat x \rangle & \langle \hat p \rangle & \cdots \\
\langle \hat x \rangle & \langle \hat x^2 \rangle & \langle \hat x \hat p \rangle & \cdots \\
\langle \hat p \rangle & \langle \hat p \hat x \rangle & \langle \hat p^2 \rangle & \cdots \\
\vdots & \vdots & \vdots & \ddots
\end{pmatrix}
\label{eq-bootstrap-matrix-anharmonic}
\end{align}
where we have taken $\{O_i\}=\{1,\hat{x},\hat{p},\cdots\}$.

We now define $E$ as the energy eigenvalue of the energy eigenstate used in the bootstrap matrix ${\mathcal M}$ \eqref{eq-bootstrap-matrix}.
Then, the energy eigenstate satisfies the following two additional conditions:
\begin{align}
	&\langle  \left[ H, O \right]   \rangle =0 ,
	\label{eq-HO=0} \\
	&\langle  OH  \rangle =E \langle  O   \rangle,
	\label{eq-HO=EO}
\end{align}
for any well-defined operator $O$, where $H$ denotes the Hamiltonian $H_{\rm AHO}$ \eqref{eq-H-anharmonic} or $H_{\rm mixed}$ \eqref{eq-H-coshx}, depending on which system we are considering.

By taking $O=\hat{x}^m \hat{p}^n $ (for the anharmonic oscillator) or $\hat{x}^l e^{m\hat{x}} \hat{p}^n $ (for the mixed potential) in the conditions \eqref{eq-HO=0} and \eqref{eq-HO=EO}, we obtain the relations among the expectation values of these operators (the details of these relations are presented in Appendix \ref{appen-sec:recursion-derivation}). These relations simplify the bootstrap matrix ${\mathcal M}$ \eqref{eq-bootstrap-matrix}.

In the case of the anharmonic oscillator, these relations lead to Eqs.~\eqref{appen-eq:HO=EO-expand}, \eqref{appen-eq:comm-HO-cosh-k=0-expand} and \eqref{appen-eq:recursion-anharmonic-xm}, and all the expectation values $ \braket{\hat{p}^l \hat{x}^m \hat{p}^n} $ appearing in the bootstrap matrix ${\mathcal M}$ \eqref{eq-bootstrap-matrix-anharmonic} can be expressed in terms of $E$, $\braket{\hat{x}}$ and $\braket{\hat{x}^2}$. Then, ${\mathcal M}$ \eqref{eq-bootstrap-matrix-anharmonic} is reduced to the following form
\begin{align}\label{anharmonic-bootstrap-matrix}
    \mathcal{M}=\begin{pmatrix}
1 & \langle \hat x \rangle & \langle \hat p \rangle & \cdots \\
\langle \hat x \rangle & \langle \hat x^2 \rangle & \langle \hat x \hat p \rangle & \cdots \\
\langle \hat p \rangle & \langle \hat p \hat x \rangle & \langle \hat p^2 \rangle & \cdots \\
\vdots & \vdots & \vdots & \ddots
\end{pmatrix}
=	\begin{pmatrix}
		1 & \left\langle \hat{x}	\right\rangle  & 0   & \cdots  \\
		\left\langle \hat{x}	\right\rangle  &  \left\langle \hat{x}^2	\right\rangle & \frac{i}{2} &  \cdots  \\
		0  &  -\frac{i}{2} & \frac{1}{3}\left( 4E- \left\langle \hat{x}^2	\right\rangle \right)  & \cdots   \\
		\vdots & \vdots  & \vdots  & \ddots \\
	\end{pmatrix},
\end{align}
and thus ${\mathcal M}$ is expressed as functions of $E$, $\braket{\hat{x}}$ and $\braket{\hat{x}^2}$. In the bootstrap method, we treat these three quantities as independent variables and ask which values satisfy the positive semidefinite condition ${\mathcal M} \succeq 0$. The region of values that satisfies ${\mathcal M} \succeq 0$ is called the allowed region, and it determines the energy eigenvalues of the system. We can ask the same question for the mixed potential to obtain its energy eigenvalues as well.
To compute the allowed region for both the anharmonic oscillator and the mixed potential, we perform numerical analysis. The details of the numerical method are described in Appendix~\ref{appen-sec:bootstrap-method-detail}.

\begin{figure}[t]
	\centering
	\includegraphics[scale=0.6]{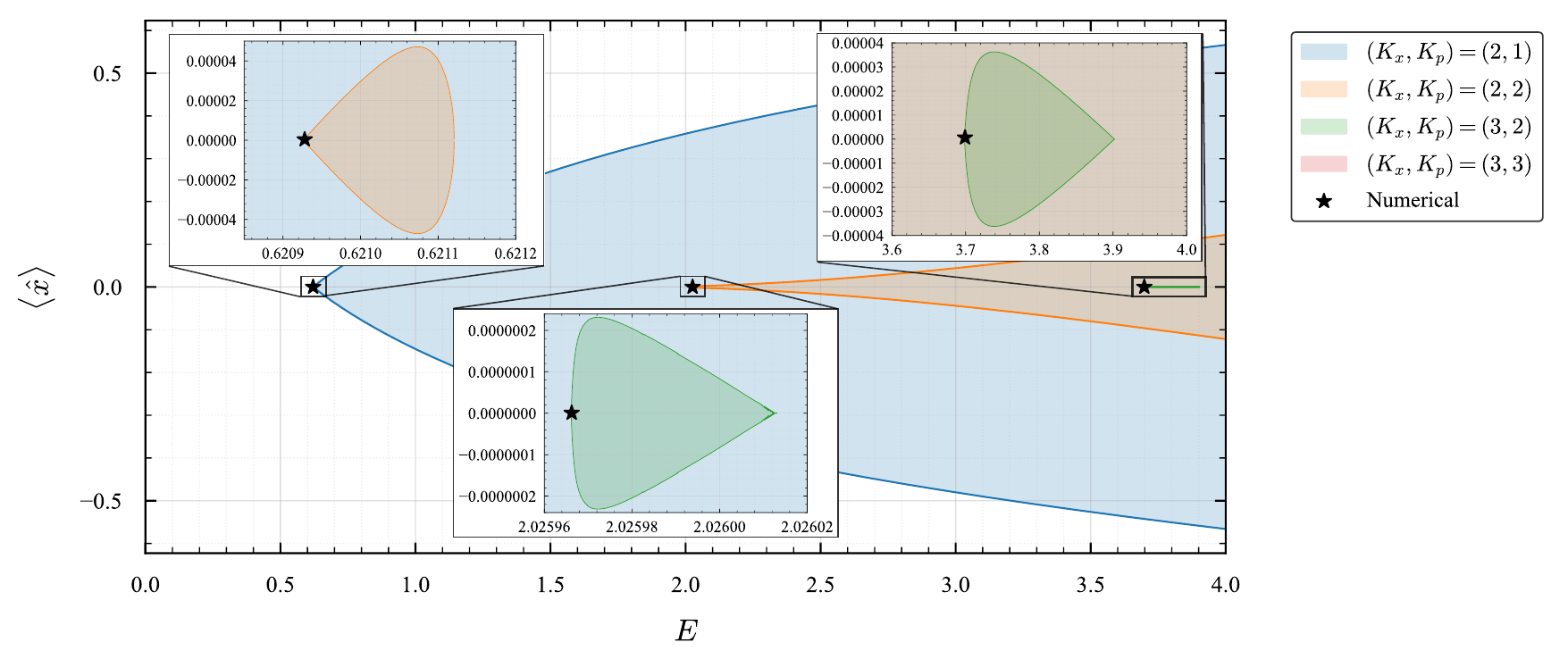}
	\caption{Numerical bootstrap result for the anharmonic oscillator \eqref{eq-H-anharmonic}.	
	The colored regions show the allowed regions on the $(E,\Braket{\hat{x}})$-plane that satisfy the condition ${\mathcal M} \succeq 0$. We use the conventional bootstrap matrix ${\mathcal M}$ with $K_x$ and $K_p$ defined in Eq.~\eqref{eq-operators-XP}.
    ``Numerical'' refers to the result obtained using the \texttt{NDEigensystem} package 
    in \textit{Mathematica}. Some of the allowed regions are difficult to see in this figure because they are extremely small, almost point-like, and their detailed values are listed in Table~\ref{table:energy-bounds-aho}.
		As $K_x$ and $K_p$ increase, the allowed regions shrink, and converge to the numerical results. 
    Here, $\Braket{\hat{x}}$ converges to 0, as expected from the parity symmetry.
	}
	\label{fig-x4-x}
\end{figure}

First, we present the results for the anharmonic oscillator. The allowed region on the $(E,\Braket{\hat{x}})$-plane  obtained through the numerical bootstrap analysis is shown in Fig.~\ref{fig-x4-x}. The detailed bounds on $E$ are summarized in Table~\ref{table:energy-bounds-aho}.
For large $K$, the allowed regions become isolated into several small regions, and as we increase the size of the bootstrap matrix $K$ (i.e., increase $K_x$ and $K_p$), these regions shrink and converge to the values of $E$ and $\braket{\hat{x}}$ at the energy eigenstates.
Therefore, for sufficiently large $K$, the energy eigenvalues of the anharmonic oscillator can be extracted from the allowed regions.
The width of each isolated allowed region can be regarded as the error in the energy eigenvalue obtained by the bootstrap method\footnote{\label{ftnt-bootstrap-error}One advantage of the bootstrap method is that the errors are ``exact'', in the sense that the true energy eigenvalues lie within the allowed regions.
(The numerically obtained allowed region is not strictly ``exact'' because of the numerical errors in the computation, but these errors are typically very small.)
Furthermore, the allowed region never enlarges as $K_x$ or $K_p$ increases. This is because the positivity condition \eqref{eq-positivity} becomes stronger as new operators are added to $\{ O_i \}$.
Thus, improving the precision (i.e., obtaining smaller allowed regions) by increasing $K_x$ and $K_p$ is theoretically guaranteed. However, obtaining the allowed regions for larger $K_x$ and $K_p$ is computationally demanding \cite{Berenstein:2022unr}, and therefore we restrict our analysis up to certain values of $K_x$ and $K_p$ in this paper.}.
Table~\ref{table:energy-bounds-aho} shows that the error is $O(10^{-8})$ for the ground state at $(K_x, K_p)=(3,3)$.
\\

\begin{table}[h]
	\centering
	\begin{tabular}{|c||c|c|c|}
		\hline
		$(K_x, K_p)$ & ground & first & second \\
		\hline
		\hline
		$(2,1)$ &  \multicolumn{3}{c|}{$0.6200<E$}  \\
		\hline
		$(2,2)$ & $0.62092600<E<0.62112068$ &  \multicolumn{2}{c|}{$2.0256<E$} \\
		\hline
		$(3,2)$ & $0.62092699<E<0.62092708$ & $2.02596609<E<2.02601309$ & $3.69840<E<3.90184$ \\
		\hline 
		$(3,3)$ & $
		0.62092702<E<0.62092706$  & $2.02596613<E<2.02596627$ & $3.698450<E<3.698454$ \\
		\hline
		Numerical & 0.620927029& 2.025966167 & 3.69845032 \\
		\hline
	\end{tabular}
	\caption{Summary of the energy bounds for the anharmonic oscillator \eqref{eq-H-anharmonic} obtained using the conventional bootstrap method at several values of $(K_x,K_p)$.
	    ``Numerical'' refers to the result obtained using the \texttt{NDEigensystem} package 
    in \textit{Mathematica}. 
		The bootstrap results converge to the numerical values as $K_x$ and $K_p$ increase.
	}
	\label{table:energy-bounds-aho}
\end{table}

\begin{figure}[htbp]
	\begin{center}
    \includegraphics[width=1.1\textwidth]{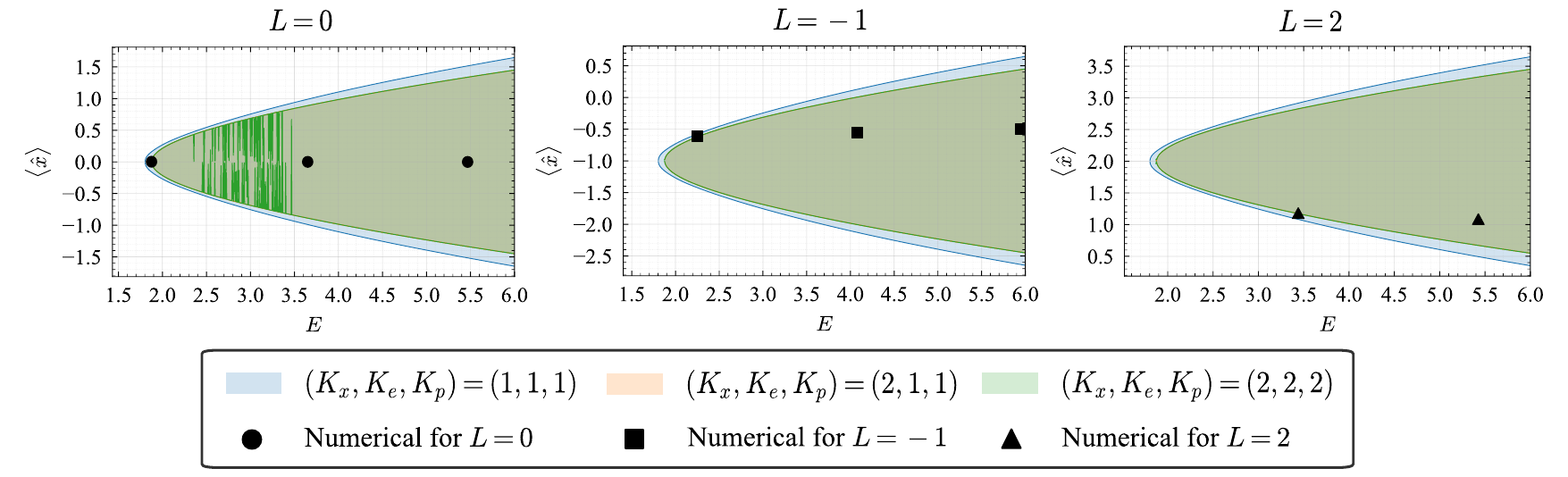}		
	\end{center}
\caption{Allowed regions for the mixed potential \eqref{eq-H-coshx} with $L=0,-1$ and $2$, obtained using the conventional bootstrap matrix. The ``numerical" points correspond to those shown in Fig.~\ref{fig-spectrum}. Up to $(K_x, K_{\rm e}, K_p) = (2, 2, 2)$, no isolated allowed regions appear.
The result for $(K_x, K_{\rm e}, K_p) = (2, 2, 2)$ at $L=0$ shows some numerical noise. These three figures become almost identical once we shift $\braket{\hat{x}} \to \braket{\hat{x}}-L$ as shown in Fig.~\ref{fig-cosh-bootstrap-shifted}, indicating that the conventional bootstrap method cannot distinguish between different values of $L$ and therefore fails to reproduce the spectrum.
    }
    \label{fig-cosh-bootstrap}
\end{figure}

Next, we present the results for the mixed potential. We consider $L=0,-1$ and $2$ as in Fig.~\ref{fig-spectrum}, and the corresponding allowed regions are shown in Fig.~\ref{fig-cosh-bootstrap}. 
We find that, up to $(K_x, K_{\rm e}, K_p) = (2, 2, 2)$, no isolated allowed regions appear for any of the three values of $L$.
Moreover, the results for $(K_x, K_{\rm e}, K_p) = (2, 1, 1)$ and $(2, 2, 2)$ are nearly identical, indicating saturation.
Therefore, it is not possible to extract the energy eigenvalues from the allowed regions. This behavior is in sharp contrast to that of the anharmonic oscillator case.

\begin{figure}[htbp]
    \centering \includegraphics[width=1.0\textwidth]{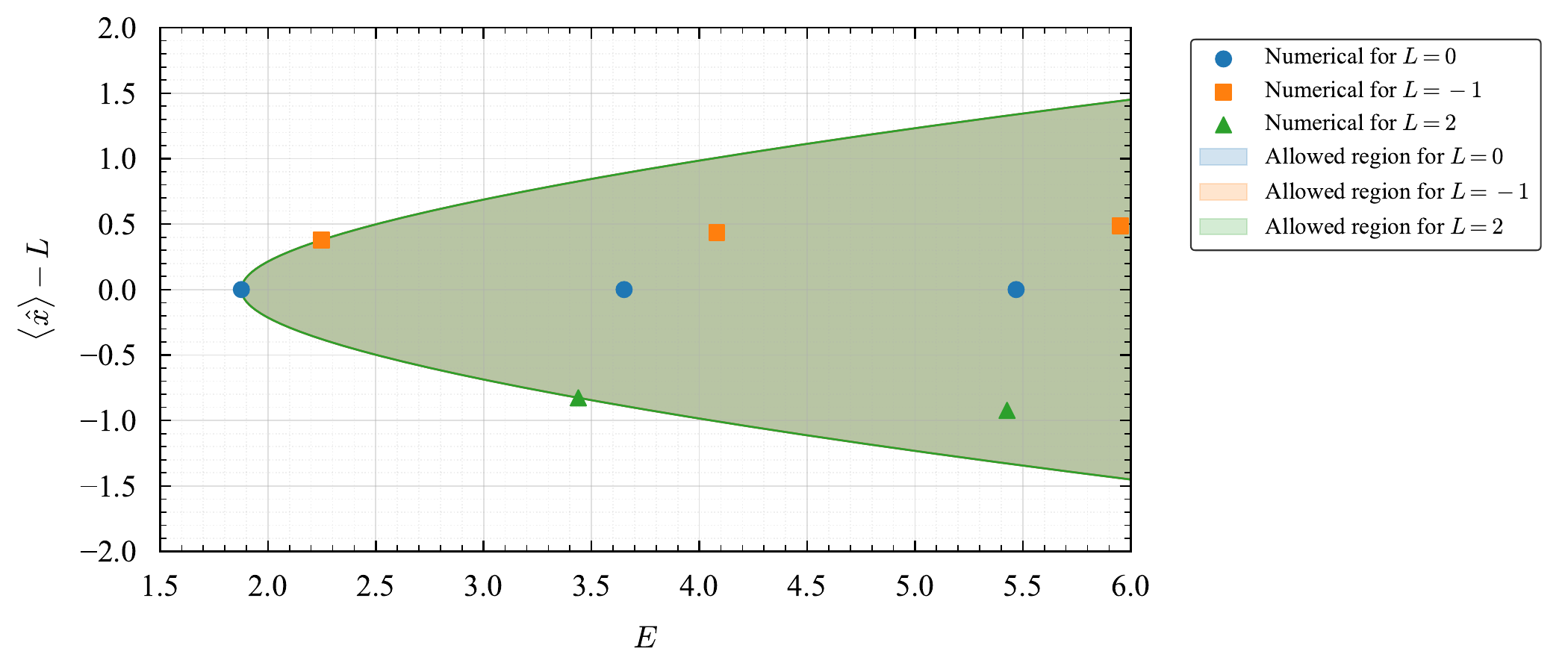}
    \caption{Allowed regions of the mixed potential \eqref{eq-H-coshx} for $L=0,-1$ and $2$ on the $(E, \braket{\hat{x}}-L)$ plane. The dots represent the numerical results shown in Fig.~\ref{fig-spectrum}. 
	We use the conventional bootstrap matrix at $(K_x, K_{\rm e}, K_p) = (2, 1, 1)$. The three allowed regions for the different values of $L$, obtained using the conventional bootstrap method and shown in Fig.~\ref{fig-cosh-bootstrap}, coincide up to numerical errors, and therefore cannot be distinguished in this plot. This demonstrates that the conventional bootstrap method cannot reproduce the energy spectrum, which depends on $L$.
    }
    \label{fig-cosh-bootstrap-shifted}
\end{figure}

Furthermore, the lower bounds of $E$ for different values of $L$ are (up to numerical errors) identical; see Table~\ref{table:energy-bounds-mixed}. In fact, if we plot $E$ against $\braket{\hat{x}}-L$ instead of $\braket{\hat{x}}$, the allowed regions for different values of $L$ completely overlap, as shown in Fig.~\ref{fig-cosh-bootstrap-shifted}. As we have seen in Fig.~\ref{fig-spectrum}, the energy eigenvalues of this system depend on the value of $L$. However, the allowed regions obtained by the bootstrap method do not depend on $L$. This implies that the implementation using $P=1$ and the operator set \eqref{eq-operators-XEP} cannot reproduce the spectrum of this system.\\

\begin{table}[h]
\centering
\begin{tabular}{|c||c|c|c|}
\hline
$(K_x,K_{\rm e},K_p)$ & $L=0$ & $L=-1$ & $L=2$ \\
\hline
\hline
$(1,1,1)$ &  $1.80311<E$ & $1.80311<E$ & $1.80311<E$ \\
\hline
$(2,1,1)$ & $1.87654<E$  & $1.87654<E$ & $1.87654<E$  \\
\hline
$(2,2,1)$ & $1.87654<E$ & $1.87654<E$ & $1.87654<E$ \\

\hline
Numerical & 1.87663935 & 2.24859879 & 3.43903964 \\
\hline
\end{tabular}
\caption{Summary of the energy bounds for the mixed potential \eqref{eq-H-coshx} with $L=0,-1,2$ at $(K_x,K_{\rm e},K_p)=(1,1,1)$, $(2,1,1)$ and $(2,2,1)$, obtained using the conventional bootstrap method. ``Numerical" refers to the ground state energies shown in Fig.~\ref{fig-spectrum}. 
The lower bounds for different values of $L$ are identical, indicating that the conventional bootstrap method cannot reproduce the spectrum of this system, which depends on $L$.
As seen in Figs.~\ref{fig-cosh-bootstrap} and \ref{fig-cosh-bootstrap-shifted}, the lower bounds are achieved at $\braket{\hat{x}}=L$.
}
\label{table:energy-bounds-mixed}
\end{table}

\subsection{Underdetermination in the conventional bootstrap method}
\label{sec-ambiguity-origin}

We have seen that the conventional bootstrap method successfully reproduces the spectrum of the anharmonic oscillator \eqref{eq-H-anharmonic}, whereas it fails to reproduce the spectrum of the mixed potential \eqref{eq-H-coshx}. Why does the conventional bootstrap method fail in the latter case? The key issue is that the conventional bootstrap method cannot fully determine the relation between the operators $\hat{x}$ and $e^{\hat{x}}$. This underdetermination effectively removes the $L$-dependence from the Hamiltonian \eqref{eq-H-coshx}, and therefore the conventional bootstrap method cannot reproduce the correct spectrum of the mixed potential.
We explain this problem in detail below.\\

In the conventional bootstrap method, we constrain the expectation values of the operators through the constraints 
	$\langle  \left[ H, O \right]   \rangle =0$ and $\langle  OH  \rangle =E \langle  O   \rangle$, and the semidefinite condition ${\mathcal M} \succeq 0$.
We also use some basic properties of the operators, such as the commutation relation $[\hat{x},\hat{p}]=i$ and the conjugation property $\braket{O^\dagger}=\braket{O}^*$. 
However, the relation between the operators $\hat{x}$ and $e^{\hat{x}}$ is not completely constrained by these conditions.

In order to see this point, we define $S := e^{\hat{x}}$.
This operator is Hermitian and satisfies the commutation relation $[\hat{p}, S]=-i S$.
However, the same relations are satisfied by the operator $S' := \pm e^{\hat{x}-x_0}$, where $x_0$ is a real constant. The other constraints used in the conventional implementation considered here also do not distinguish $S$ from $S'$.
(If we regard $[\hat{p}, S']=-i S'$ as a differential equation for $S'$, the conventional constraints do not specify its initial condition. Consequently, the constant factor $\pm e^{-x_0}$ remains undetermined.)

Usually the operators $\hat{x}$ and $e^{\hat{x}}$ are related through the Taylor expansion
\begin{align} 
  e^{\hat{x}}=\sum_{n=0}^\infty \frac{1}{n!} (\hat{x})^n ,
 \end{align}
but the conventional bootstrap method does not use such a relation.
Thus, the operators $\hat{x}$ and $e^{\hat{x}}$ are not directly related in this sense, and the factor $\pm e^{-x_0} $ in $S'$ is undetermined.
Similarly, the operator $S^m =  e^{m\hat{x}}$ is not distinguished from the operator $S'^m = (\pm)^m e^{m(\hat{x}-x_0)}$ for any integer $m$.

Now we discuss the consequences of this underdetermination when we derive the allowed region of the mixed potential.
Since the conventional constraints do not distinguish $e^{\hat{x}}$ from $ \pm e^{\hat{x}-x_0}$, they are effectively compatible with the following replacement:
\begin{align}
    H_{\rm mixed} & = \frac{1}{2}\hat{p}^2 + (\hat{x}-L)^2 +\cosh(\hat{x}) \quad  \to \quad H'_{\rm mixed} := \frac{1}{2}\hat{p}^2 + (\hat{x}-L)^2  \pm \cosh(\hat{x}-x_0)  .
\label{eq-H-mixed-ambiguity}
\end{align}

\begin{figure}[htbp]
    \centering
    \includegraphics[width=\textwidth]{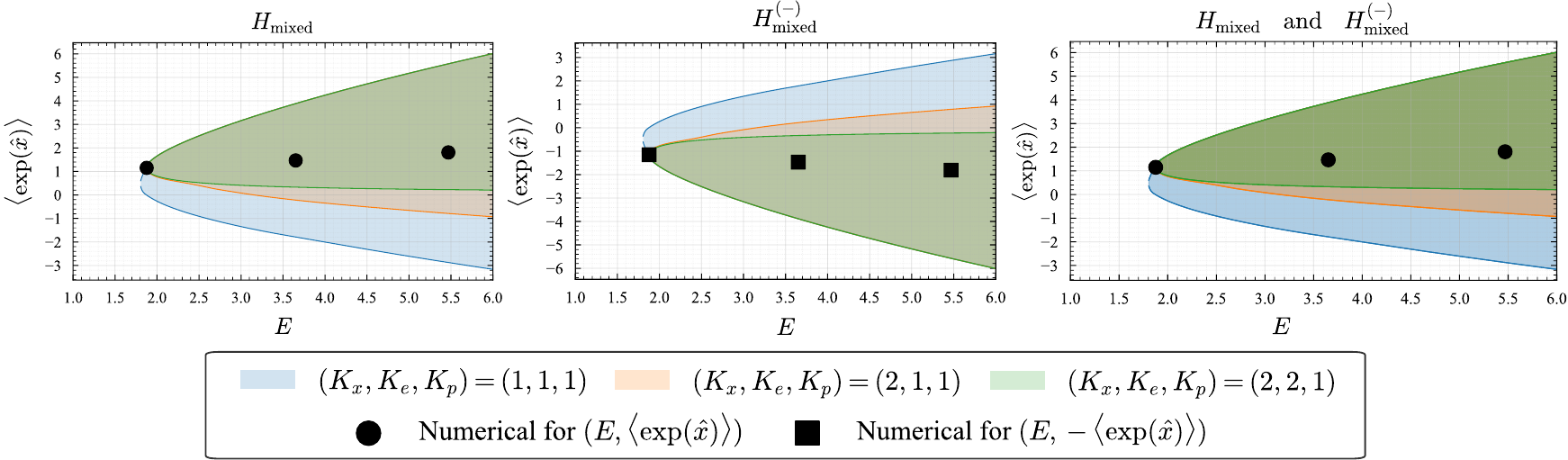}
 \caption{ The allowed regions for $H_{\rm mixed}$ in Eq.~\eqref{eq-H-coshx} and $H^{(-)}_{\rm mixed}$ in Eq.~\eqref{eq-H-mixed-minus} at $L=0$, obtained using the conventional bootstrap method. Left: the allowed regions in the $(E,\braket{e^{\hat{x}}})$ plane for $H_{\rm mixed}$. Center: the allowed regions in the $(E,\braket{e^{\hat{x}}})$ plane for $H^{(-)}_{\rm mixed}$. The black square markers represent $(E,-\braket{e^{\hat{x}}})$ for the eigenstates of $H_{\rm mixed}$, not of $H^{(-)}_{\rm mixed}$. Right: an overlay of the left and center panels, with the center panel reflected across the $\braket{e^{\hat{x}}}=0$ axis. After this reflection, the allowed regions for $H_{\rm mixed}$ and $H^{(-)}_{\rm mixed}$ coincide, demonstrating the sign underdetermination of $e^{\hat{x}}$ in the conventional bootstrap method. }
    \label{fig-cosh-bootstrap-minus}
\end{figure}

First, we consider the underdetermination of the $\pm$ sign in front of the $\cosh(\hat{x}-x_0)$ term. To confirm the existence of this underdetermination, we introduce the Hamiltonian
\begin{align}
    H^{(-)}_{\rm mixed} &:= \frac{1}{2}\hat{p}^2 + (\hat{x}-L)^2 -\cosh(\hat{x}) ,
\label{eq-H-mixed-minus}
\end{align}
in which the sign of the $\cosh(\hat{x})$ term is flipped from the original $ H_{\rm mixed}$.
If the sign underdetermination is present in the conventional bootstrap method, the allowed regions for $H^{(-)}_{\rm mixed}$ should be related to those for $H_{\rm mixed}$ by $\braket{e^{\hat{x}}} \to -\braket{e^{\hat{x}}}$.
We perform the conventional bootstrap analysis for this Hamiltonian and find that the two allowed regions are indeed related in this way, as shown in Fig.~\ref{fig-cosh-bootstrap-minus}.
This result indicates the following two points:
\begin{enumerate}
	\item The conventional bootstrap method cannot distinguish between the operators $e^{\hat{x}}$ and $-e^{\hat{x}}$, as we expected.
	\item If the minus sign in $H_{\rm mixed}$ \eqref{eq-H-mixed-ambiguity} were taken, the potential would be unbounded from below, and no lower bound on $E$ would be expected. However, no such contribution is observed in the obtained allowed region. Similarly, no contribution from the positive-sign branch is observed in the allowed region for $H^{(-)}_{\rm mixed}$.
\end{enumerate}
The reason for the second point is unclear, but it may be related to the fact that the energy eigenstates of the unbounded Hamiltonian are not normalizable, and the expectation values of the operators may not be well-defined.
To understand this point, we should clarify the properties of the bootstrap method for unbounded Hamiltonians, but this issue is beyond the scope of this paper, and we leave it for future work\footnote{
Unlike the underdetermination of $x_0$, the sign underdetermination arises even when the Hamiltonian does not contain a polynomial potential. For example, the model $H=\hat{p}^2/2+\cosh(\hat{x})$ was studied in Ref.~\cite{Du:2021hfw}, where its spectrum was successfully reproduced using the conventional bootstrap method. We have also confirmed that, under the reflection $\braket{e^{\hat{x}}}\to-\braket{e^{\hat{x}}}$, the allowed region for $H=\hat{p}^2/2-\cosh(\hat{x})$ is mapped onto that for $H=\hat{p}^2/2+\cosh(\hat{x})$, in the same manner as in Fig.~\ref{fig-cosh-bootstrap-minus}.}.

Now, we consider the underdetermination of $x_0$ in $H'_{\rm mixed} $ \eqref{eq-H-mixed-ambiguity}.
Because of this underdetermination, the conventional bootstrap method may yield an allowed region that satisfies the condition ${\mathcal M} \succeq 0$ for arbitrary values of $x_0$.
Thus it effectively removes the $L$-dependence from the Hamiltonian \eqref{eq-H-coshx}.
This explains why the allowed region for $E$ obtained by the conventional bootstrap method does not depend on $L$, as shown in Table~\ref{table:energy-bounds-mixed} and Fig.~\ref{fig-cosh-bootstrap-shifted}.

Such underdetermination may arise when the conventional bootstrap method is applied to a system involving functional operators whose mutual relations are not fully determined by the constraints used.
This could pose a general problem for the conventional bootstrap method\footnote{See Ref.~\cite{Lawrence:2025wyl} for a related potential problem and Refs.~\cite{Aikawa:2021eai, Berenstein:2021loy,Tchoumakov:2021mnh, Huang:2025sua} for underdetermination of $\hat{p}$. 
Similarly, if we evaluate the expectation value of an operator that is not related to the operators appearing in the Hamiltonian, the underdetermination may occur and the conventional bootstrap method may fail to reproduce the expectation value. 
For example, evaluating $\braket{e^{\hat{x}}}$ for the energy eigenstates of the anharmonic oscillator \eqref{eq-H-anharmonic} would be difficult \cite{Morita:WIP}. 
See Ref.~\cite{Aikawa:2025dvt} for a related problem in the hyperbolic Scarf potential.
}.

\section{Possible resolutions to the underdetermination}
\label{sec-ambiguity-resolution}

We have seen that the conventional bootstrap method cannot distinguish the operator $e^{\hat{x}}$ from $\pm e^{\hat{x}-x_0}$. In this section, we discuss possible resolutions to this underdetermination.

Since the underdetermination of the $\pm$ sign may not be the main obstacle to obtaining the spectrum of $H_{\rm mixed}$, we focus on the underdetermination of $x_0$ in this section. (We discuss resolutions to the sign underdetermination in Appendix \ref{appen-sign}.)

We will propose three possible prescriptions.
In particular, we will show that the third resolution yields the isolated allowed regions that converge to the energy eigenvalues, and thus it is a promising resolution to the underdetermination.
These resolutions may also be applicable to systems in which the underdetermination leads to difficulties.

\subsection{Using Taylor expansion}
\label{sec-taylor-expansion}
If the cosh function is expanded in a Taylor series and truncated at a finite order $n_{\rm max}$ as
\begin{align} 
\cosh(\hat{x}) = \sum_{n=0}^\infty \frac{(\hat{x})^{2n}}{(2n)!} \ \to \ \sum_{n=0}^{n_{\rm max}} \frac{(\hat{x})^{2n}}{(2n)!} ,
\label{eq-cosh-Taylor}
\end{align}
the underdetermination does not occur, since the potential involves only polynomial operators $(\hat{x})^n$.
(We have expanded $\cosh(x)$ around $x = 0$, but it can be expanded around an arbitrary point.)

However, this prescription spoils one of the key strengths of the bootstrap method: the allowed region is ``exact''.
Specifically, the energy eigenvalue never takes a value outside the allowed region (see footnote \ref{ftnt-bootstrap-error} as well).
This property is lost in this prescription because, as long as $n_{\rm max}$ is finite, $\cosh(\hat{x})$ and its truncated version in Eq.~\eqref{eq-cosh-Taylor} differ, and the energy spectra for these two potentials do not agree.
Thus, even if we perform the bootstrap analysis using the prescription \eqref{eq-cosh-Taylor}, the true energy eigenvalues may lie outside the obtained allowed region.
Therefore, the allowed region cannot be exact in this prescription.

Because of this issue, we do not pursue this prescription in this article.

\subsection{Imposing parity condition}
\label{sec-parity-condition}
In the case of $L=0$ in the mixed potential \eqref{eq-H-coshx}, the Hamiltonian is invariant under the parity transformation $\hat{x} \to -\hat{x}$ and $\hat{p} \to -\hat{p}$.
This symmetry may resolve the underdetermination in this case.

The parity symmetry requires that the energy eigenstates are either parity even or parity odd.
Thus, the expectation values of the operators at the energy eigenstates satisfy
\begin{align}
   \braket{ \hat{x}^l e^{m\hat{x}} \hat{p}^n} = (-1)^{l+n}  \braket{ \hat{x}^l e^{-m\hat{x}} \hat{p}^n} .
   \label{eq-parity-condition}
\end{align}
Because of the underdetermination of $e^{\hat{x}}$, the conventional bootstrap method may regard the parity transformation as $ \pm e^{\hat{x}-x_0} \to \pm e^{-\hat{x}-x_0}$.
This operation is consistent with the parity condition \eqref{eq-parity-condition} only when $x_0=0$.
Thus, if we add the condition \eqref{eq-parity-condition} to the constraints \eqref{eq-HO=0} and \eqref{eq-HO=EO} used in the conventional bootstrap method, the underdetermination may be resolved.

\begin{figure}[htbp]
    \centering
    \includegraphics[width=\textwidth]{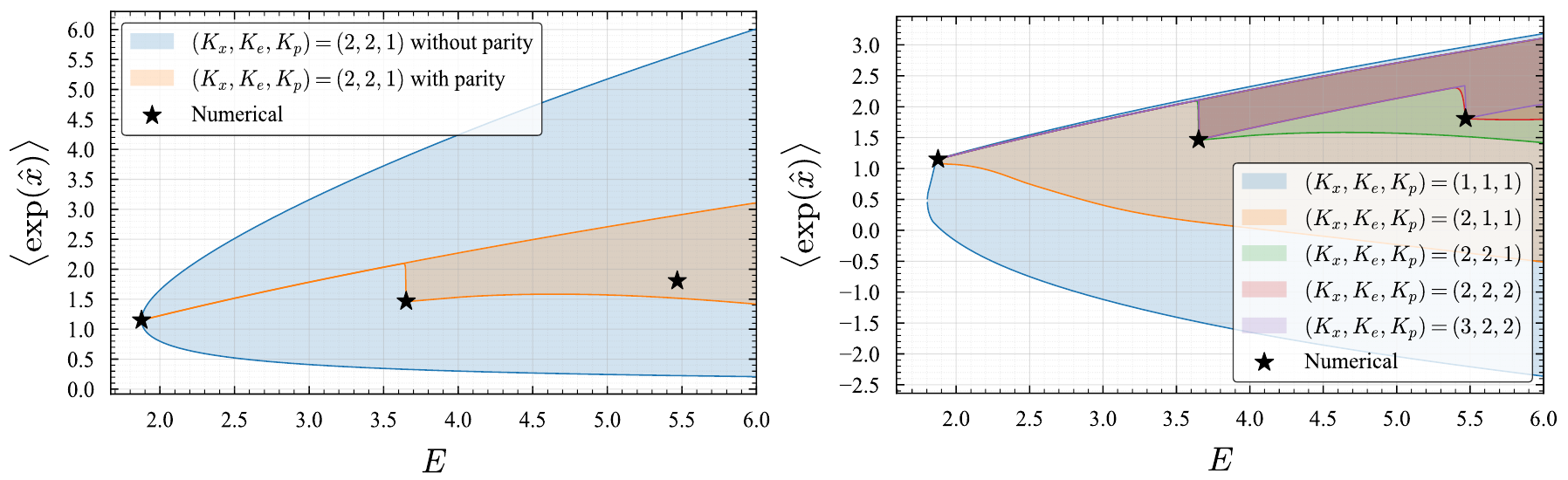}
    \caption{The allowed regions of the mixed potential \eqref{eq-H-coshx} for $L=0$ obtained using the conventional bootstrap method with the parity condition \eqref{eq-parity-condition}.
    The vertical axis is $\braket{ {\rm exp}(\hat{x})} $ instead of $\braket{\hat{x}}$, because the condition $\braket{\hat{x}}=0$ has been imposed. 
	Left: comparison of the allowed region with and without imposing the parity condition at $(K_x, K_{\rm e}, K_p) = (2, 2, 1)$. 
	The allowed region is drastically reduced by imposing the parity condition. Right: results at various $(K_x, K_{\rm e}, K_p)$. As we increase the size of the conventional bootstrap matrix, the allowed region shrinks and kinks appear. The energy eigenvalues are located at these kinks, indicating that the bootstrap analysis captures the eigenstates.
	Although isolated allowed regions have not been obtained at these bootstrap matrix sizes, it appears that the underdetermination is resolved by the parity condition \eqref{eq-parity-condition}.
    }
    \label{fig-cosh-L=0-bootstrap}
\end{figure}

Let us check whether this prescription works in the mixed potential \eqref{eq-H-coshx} at $L=0$.
We impose the parity condition \eqref{eq-parity-condition} on the operators used in the numerical bootstrap analysis (the details of the analysis are given in Appendix \ref{appen-sec:parity}).
The obtained allowed region in the $(E,\Braket{e^{\hat{x}}})$-plane is shown in Fig.~\ref{fig-cosh-L=0-bootstrap}.
We see that kinks appear in the allowed region, and the energy eigenvalues are located at these kinks. Thus, the bootstrap method captures the energy eigenstates, although isolated allowed regions have not been obtained at these bootstrap matrix sizes. 
This suggests that the parity condition \eqref{eq-parity-condition} resolves the underdetermination. We expect that, if we increase the size of the bootstrap matrix further, isolated allowed regions will appear and converge to the energy eigenvalues, although this would be computationally challenging. (As we will see in the next subsection, isolated allowed regions are obtained by combining another prescription, and we do not pursue obtaining isolated allowed regions solely by using the parity condition in this article.)

However, this prescription is applicable only when the system possesses parity symmetry, and therefore we need to find other prescriptions for cases in which the parity symmetry is absent.

\subsection{Imposing additional positivity condition}
\label{sec-additional-inequalities}

The cosh function satisfies the inequality
\begin{align}
    \cosh(x) \ge  \sum_{n=0}^{N_{\rm max}} \frac{x^{2n}}{(2n)!},
    \label{eq-cosh-inequality}
\end{align}
where $N_{\rm max}$ is a non-negative integer.
This inequality may be used to resolve the underdetermination.

Using the positive operator defined by this inequality as $P$ in Eq.~\eqref{eq-positivity}, we construct a new bootstrap matrix,
\begin{align} 
  {\mathcal M}^{\cosh}_{ij} = \braket{
O_i^\dagger
\left(
\cosh(\hat{x}) -  \sum_{n=0}^{N_{\rm max}} \frac{\hat{x}^{2n}}{(2n)!}
\right)
O_j
}, \quad {\mathcal M}^{\cosh}  \succeq 0 .
    \label{eq-bootstrap-matrix-cosh-positive}
\end{align}
Now, because of the underdetermination, the bootstrap method may effectively replace this condition as
\begin{align} 
  {\mathcal M}^{\cosh}_{ij} \to {\mathcal M'}^{\cosh}_{ij}= \braket{
{O'_i}^{\dagger}
\left(
\pm \cosh(\hat{x}-x_0) -  \sum_{n=0}^{N_{\rm max}} \frac{\hat{x}^{2n}}{(2n)!}
\right)
O'_j
}, \quad {\mathcal M'}^{\cosh}  \succeq 0 ,
    \label{eq-bootstrap-matrix-cosh-positive-2}
\end{align}
where the operator $O'_i$ is the one obtained by replacing $e^{\hat{x}}$ in $O_i$ with $ \pm e^{\hat{x}-x_0}$.
In the limit $N_{\rm max} \to \infty$, this semidefinite condition holds only for $x_0=0$ and the positive sign.
Thus, at finite $N_{\rm max}$, this semidefinite condition favors $x_0 = 0 $ and the positive sign and may therefore resolve both underdeterminations.

One might think that this prescription is similar to the Taylor expansion prescription \eqref{eq-cosh-Taylor} discussed in Section~\ref{sec-taylor-expansion}. 
However, these two prescriptions are in fact very different. The crucial point is that, in the present prescription, the Hamiltonian is not approximated by the Taylor expansion \eqref{eq-cosh-Taylor}. Therefore, the allowed region remains ``exact''. This is an advantage of this prescription.

\begin{figure}[htbp]
    \centering
     \includegraphics[width=\textwidth]{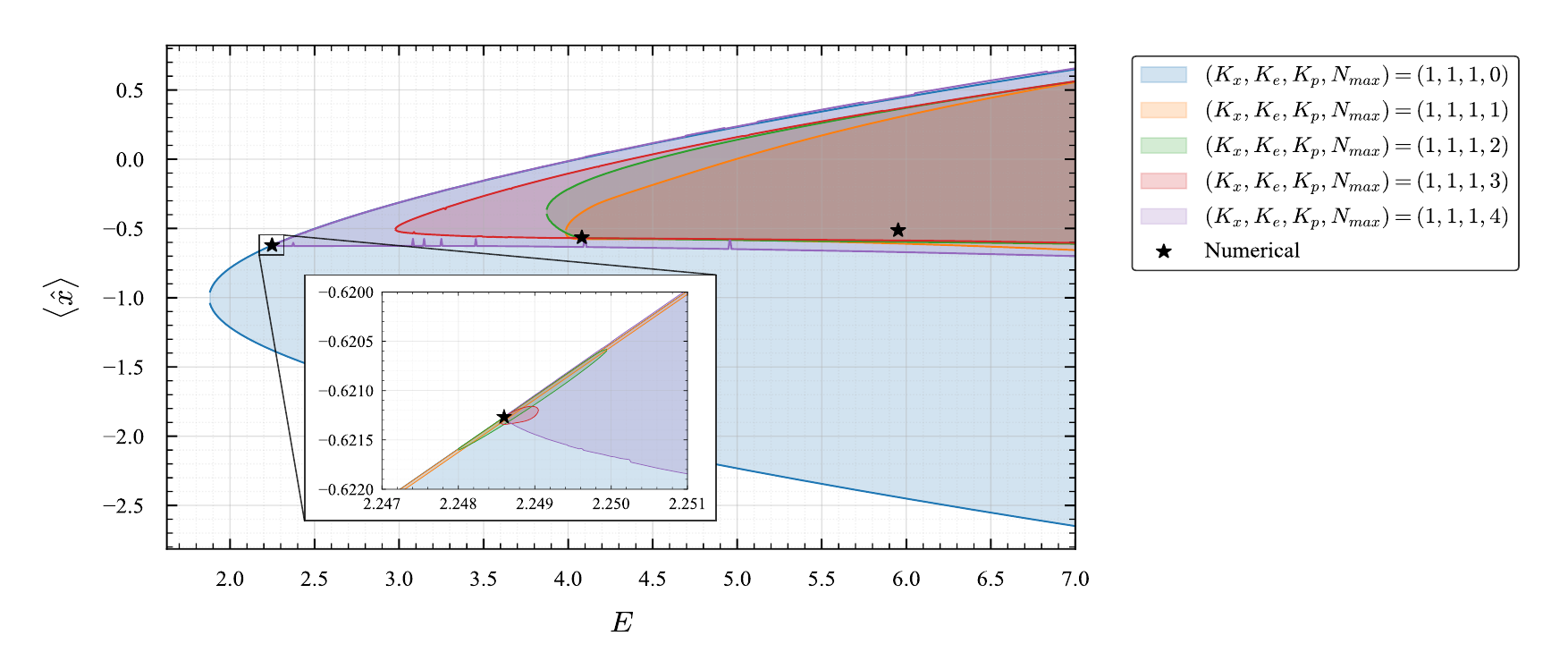}
	\includegraphics[width=\textwidth]{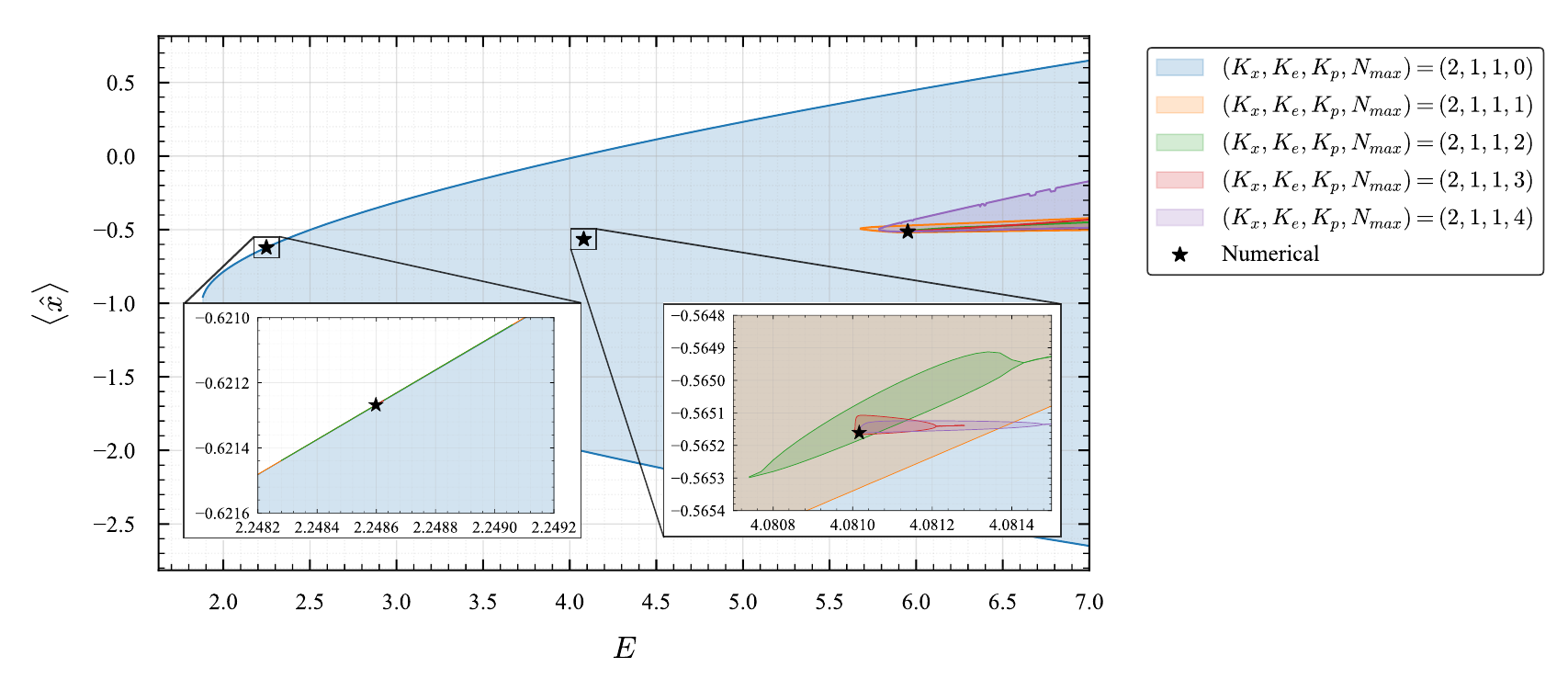}
	 \caption{Allowed regions of the mixed potential \eqref{eq-H-coshx} for $L=-1$ obtained using the bootstrap method with the additional positivity condition  ${\mathcal M}^{\cosh}  \succeq 0 $ \eqref{eq-bootstrap-matrix-cosh-positive}. The insets show magnified views around the eigenvalues.
	 The bootstrap matrices ${\mathcal M}$ and $  {\mathcal M}^{\cosh}$ are constructed from the operators $\{\hat{x}^k  e^{m\hat{x}} \hat{p}^n   \}$, ($k=0,1, \cdots, K_x$, $m=-K_{\rm e}, \cdots, K_{\rm e}$, $n=0,1, \cdots, K_p)$, where the same values of $(K_x, K_{\rm e}, K_p)$ are used for both matrices.
The underdetermination is resolved, and isolated allowed regions reproducing the energy spectrum are obtained. 
The detailed numerical bounds are summarized in Table~\ref{table:energy-bounds-add-matrix}.
    }
    \label{fig-cosh-L=-1-bootstrap}
\end{figure}

Let us see how this prescription works in the mixed potential \eqref{eq-H-coshx}. 
We perform the numerical bootstrap analysis by imposing the positivity condition \eqref{eq-bootstrap-matrix-cosh-positive} in addition to the conventional positivity condition ${\mathcal M} \succeq 0$. The details of the numerical method are described in Appendix \ref{appen-sec:add-pos-condition}.
The obtained allowed regions at $L=-1$ in Eq.~\eqref{eq-H-coshx} are shown in Fig.~\ref{fig-cosh-L=-1-bootstrap}, and the detailed data are summarized in Table~\ref{table:energy-bounds-add-matrix}. We obtain isolated allowed regions that converge to the energy eigenvalues, showing that the bootstrap method successfully reproduces the spectrum. The error of the ground state at $(K_x, K_{\rm e}, K_p, N_{\rm max})=(2,1,1,4)$ is $O(10^{-6})$.

One feature of this result is that a larger $N_{\rm max}$ does not always give a better result.
A part of the allowed region that is excluded at smaller $N_{\rm max}$ is included in the allowed region at larger $N_{\rm max}$\footnote{The allowed region always shrinks as $K_x$, $K_{\rm e}$ or $K_p$ in Eq.~\eqref{eq-operators-XEP} increases, as discussed in footnote~\ref{ftnt-bootstrap-error}. On the other hand, it is not guaranteed that the allowed region shrinks as $N_{\rm max}$ increases.}. 
We speculate that this is because the number of independent variables involved in the matrix ${\mathcal M}^{\cosh}$ increases as $N_{\rm max}$ increases, while the size of the matrix remains the same.
For example, for $(K_x,K_{\rm e},K_p)=(1, 1, 1)$ where the size of the bootstrap matrix  ${\mathcal M}^{\cosh} $ is $12 \times 12$, the numbers of independent variables involved in ${\mathcal M}^{\cosh}$ are 19, 23, 27 and 31 for $N_{\rm max}=1$, $2$, $3$ and $4$, respectively.
Since more independent variables may relax the positivity conditions \eqref{eq-bootstrap-matrix-cosh-positive}, a larger $N_{\rm max}$ does not always give a better result.

\begin{figure}[htbp]
	\centering
	\includegraphics[width=\textwidth]{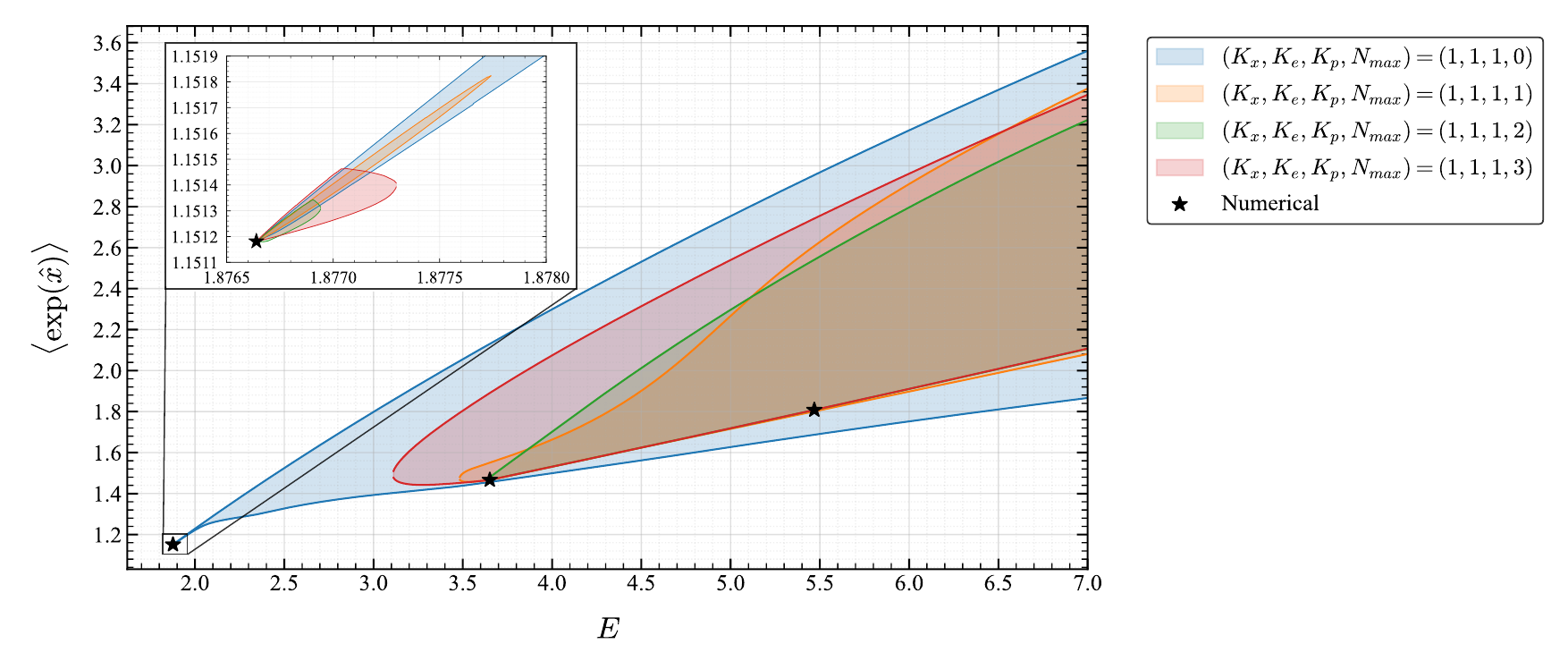}
	\includegraphics[width=\textwidth]{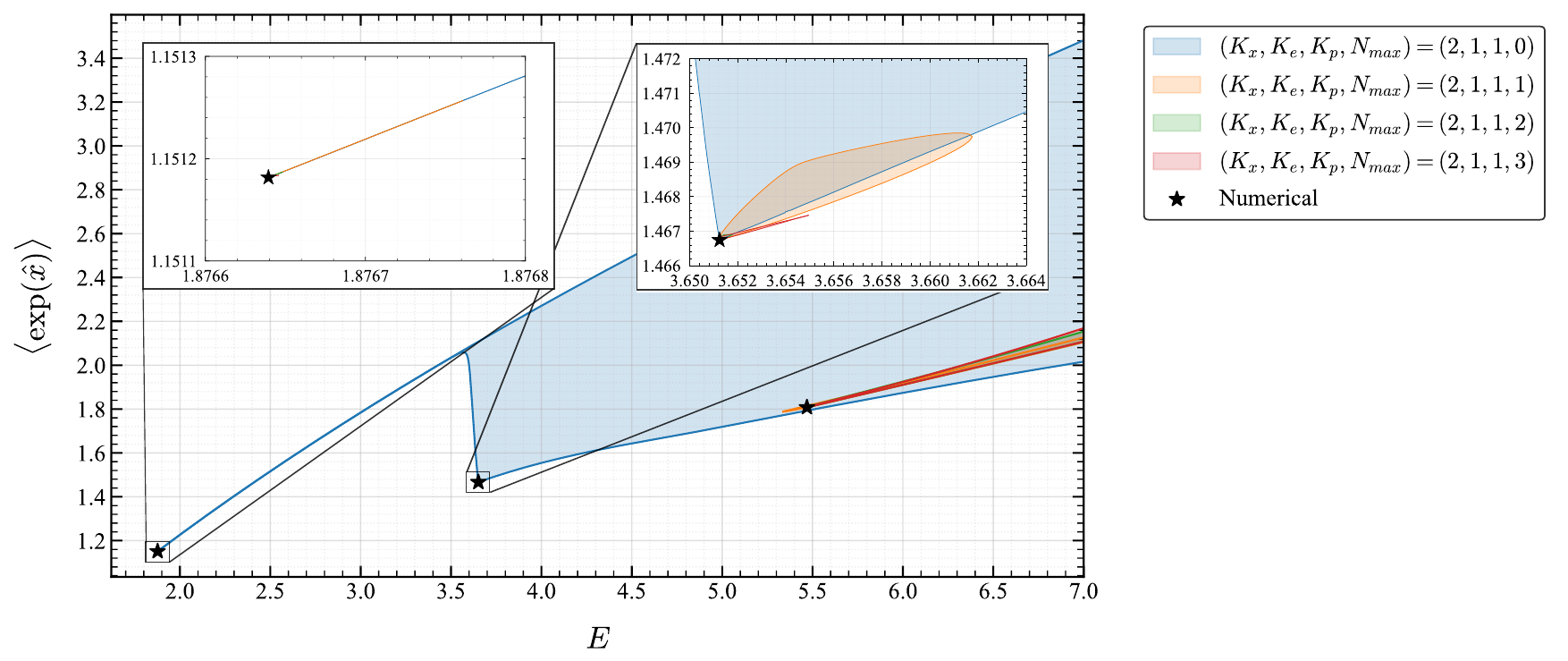}
	\caption{Allowed regions of the mixed potential \eqref{eq-H-coshx} for $L=0$, obtained using the bootstrap method with both the parity condition \eqref{eq-parity-condition} and the additional positivity condition ${\mathcal M}^{\cosh}  \succeq 0 $  \eqref{eq-bootstrap-matrix-cosh-positive}.
	The bootstrap matrices ${\mathcal M}$ and $  {\mathcal M}^{\cosh}$ are constructed from the operators $\{\hat{x}^k  e^{m\hat{x}} \hat{p}^n   \}$, $(k=0,1, \cdots, K_x$, $m=-K_{\rm e}, \cdots, K_{\rm e}$, $n=0,1, \cdots, K_p)$, using common values of $(K_x, K_{\rm e}, K_p)$ for both matrices.
	The underdetermination is resolved, and isolated allowed regions are obtained. 
	The detailed numerical bounds are summarized in Table~\ref{table:energy-bounds-add-matrix-parity}.
	}
	\label{fig:cosh-2mat-L=0-bootstrap}
\end{figure}

For the $L=0$ case, we can impose both the parity condition \eqref{eq-parity-condition} and the additional positivity condition \eqref{eq-bootstrap-matrix-cosh-positive}.
The obtained allowed region is shown in Fig.~\ref{fig:cosh-2mat-L=0-bootstrap} and Table~\ref{table:energy-bounds-add-matrix-parity}.
The results are much better than those obtained by imposing only the parity condition \eqref{eq-parity-condition}, shown in Fig.~\ref{fig-cosh-L=0-bootstrap}, and isolated allowed regions are obtained at smaller bootstrap matrix sizes. Thus, again, our prescription overcomes the underdetermination problem. The error at $(K_x, K_{\rm e}, K_p)=(2,1,1)$ is $O(10^{-5})$ for the ground state.\\

It is straightforward to apply this prescription to general cases, where the system contains multiple types of functions, if there are inequalities similar to \eqref{eq-cosh-inequality} for these functions.\\

\begin{table}[h]
	\centering
	\begin{tabular}{|c||c|c|}
		\hline
		$(K_x, K_{\rm e}, K_p, N_{\rm max})$ & ground & first \\
		\hline
		\hline
		$(1,1,1,0)$ & \multicolumn{2}{c|}{$1.876<E$}   \\
		\hline
		$(1,1,1,1)$ & $2.23520<E<2.26584$ & $3.982 < E $\\
		\hline
		$(1,1,1,2)$ & $2.24798<E<2.24996$ & $3.868 < E $\\
		\hline 
		$(1,1,1,3)$ & $2.248544<E<2.249056 $ & $2.974 < E $ \\
		\hline 
		$(1,1,1,4)$ & \multicolumn{2}{c|}{$2.24858<E$}   \\
		\hline
		\hline
		$(2,1,1,0)$ &  \multicolumn{2}{c|}{$1.876<E$}  \\
		\hline
		$(2,1,1,1)$ & $2.23921<E<2.26000$ & $4.07198<E<4.09199$  \\
		\hline
		$(2,1,1,2)$ & $2.24826<E<2.24908$ & $4.08071<E<4.08161$  \\
		\hline 
		$(2,1,1,3)$ & $2.24858798<E<2.24862488$ & $4.0810032<E<4.0812816$ \\
		\hline
		$(2,1,1,4)$ & $2.24859806<E<2.24860106$ & $4.0810152<E<4.0814992$ \\
		\hline
		Numerical & 2.24859879 & 4.08101643 \\
		\hline
	\end{tabular}
	\caption{Summary of the energy bounds of the mixed potential \eqref{eq-H-coshx} for $L=-1$, obtained using the bootstrap method with the additional positivity condition  ${\mathcal M}^{\cosh}  \succeq 0 $ \eqref{eq-bootstrap-matrix-cosh-positive}.
	The corresponding allowed regions are shown in Fig.~\ref{fig-cosh-L=-1-bootstrap}.
	Isolated allowed regions that reproduce the energy spectrum are obtained.
	Note that a larger value of $N_{\rm max}$ does not always yield a better result.
	}
	\label{table:energy-bounds-add-matrix}
\end{table}

\begin{table}[h]
	\centering
	\begin{tabular}{|c||c|c|}
		\hline
		$(K_x, K_{\rm e}, K_p,N_\text{max})$ & ground & first \\
		\hline
		\hline
		$(1,1,1,0)$ & \multicolumn{2}{c|}{$1.8766384<E$} \\
		\hline
		$(1,1,1,1)$ & $1.876638<E<1.877743$ & $3.478<E$   \\
		\hline
		$(1,1,1,2)$ & $1.876638<E<1.876944$ & $3.634<E$  \\
		\hline 
		$(1,1,1,3)$ & $1.876638<E<1.877299$ & $3.106<E $ \\
		\hline
		\hline
		$(2,1,1,0)$ & \multicolumn{2}{c|}{$1.876638<E$} \\
		\hline
		$(2,1,1,1)$ & $1.876638<E<1.876764$ & $3.65120<E<3.66177$   \\
		\hline
		$(2,1,1,2)$ & $1.876638<E<1.876651$ & $3.65123<E<3.65180$  \\
		\hline 
		$(2,1,1,3)$ & $1.876638<E<1.876648$ & $3.65123<E<3.65497$ \\
		\hline
		Numerical & 1.87663935 & 3.65124961 \\
		\hline
	\end{tabular}
	\caption{Summary of the energy bounds of the mixed potential \eqref{eq-H-coshx} for $L=0$ obtained using the bootstrap method with the parity condition \eqref{eq-parity-condition} and the additional positivity condition  ${\mathcal M}^{\cosh}  \succeq 0 $  \eqref{eq-bootstrap-matrix-cosh-positive}.
	The corresponding allowed regions are shown in Fig.~\ref{fig:cosh-2mat-L=0-bootstrap}.
	}
	\label{table:energy-bounds-add-matrix-parity}
\end{table}

\section{Discussion}
\label{sec-discussion}

In this article, we have shown that the conventional bootstrap method cannot reproduce the spectrum of the Hamiltonian with the mixed potential \eqref{eq-H-coshx}. 

We have identified the origin of this failure as an underdetermination of the relations between different operator sectors, and we have proposed three possible resolutions to this problem.
In particular, the prescription based on imposing the additional positivity condition \eqref{eq-bootstrap-matrix-cosh-positive} is promising, and it successfully reproduces the spectrum of the mixed potential \eqref{eq-H-coshx}.

This underdetermination can be a serious issue in conventional bootstrap analyses involving different functional operator sectors in the action or observables, and it may arise in various systems, including statistical models and matrix models. Although our proposal resolves the underdetermination for the mixed potential \eqref{eq-H-coshx}, it is not clear whether it always works in more general settings.
Since the bootstrap method is a powerful tool for analyzing such systems, it is important to develop a general prescription that can systematically resolve the underdetermination.\\

Furthermore, the allowed regions obtained for the mixed potential, shown in Tables~\ref{table:energy-bounds-add-matrix} and \ref{table:energy-bounds-add-matrix-parity}, are not as sharp as those for the anharmonic oscillator in Table~\ref{table:energy-bounds-aho}.
Performing the bootstrap analysis with larger values of $K_x$, $K_{\rm e}$ and $K_p$ may improve the results, but this poses a significant numerical challenge.

\paragraph*{Acknowledgment.---}
The authors would like to thank Y.~Hashimoto for collaboration in the early stages of this project. T.M. and P.S. gratefully acknowledge the International Centre for Theoretical Sciences (ICTS) for the opportunity to participate in the Hydrodynamics, Fluctuations, and Noise in Quantum and Classical Systems (code: ICTS/hydrodynamics2025/12), where this collaboration was initiated. The work of T.~M. is supported in part by Grant-in-Aid for Scientific Research C (No. 20K03946) from JSPS. P.S. is supported by an IIT Kanpur Institute Assistantship.

\paragraph*{Data availability.---}
There are no publicly available research data or software supporting this manuscript. Requests for further information or data should be sent to the authors.

\appendix

\section{Derivation of recursion relations} 
\label{appen-sec:recursion-derivation}

In this appendix, we apply the constraint relations \eqref{eq-HO=0} and \eqref{eq-HO=EO} to the expectation values of the operators in a systematic manner. 
This leads to recursion relations among the expectation values, allowing us to express them in terms of a much smaller set of quantities, which we refer to as independent variables. 
These relations simplify the numerical bootstrap analysis presented in the main text.

\subsection{General structure of the constraint equations \eqref{eq-HO=0} and \eqref{eq-HO=EO}} 
\label{appen-sec:recursion-derivation-general}

In this subsection, we consider the Hamiltonian
\begin{align}
    H = \frac{1}{2}\hat{p}^2 + V(\hat{x}) \ ,
    \label{eq-Hamiltonian-general}
\end{align}
where we do not specify the explicit form of the potential $V(\hat{x})$, and discuss the general structure of the recursion relations obtained from the constraint equations $\langle [H,O] \rangle = 0$  \eqref{eq-HO=0} and $\langle OH \rangle = E \langle O \rangle$ \eqref{eq-HO=EO}.

Note that, by using the commutation relation $[\hat{p},\hat{x}] = -i$, any operator of the form $\hat{p}^n O(\hat{x}) $, where $O(\hat{x})$ is a function of $\hat{x}$, can be rearranged as
\begin{align} 
  \hat{p}^n O(\hat{x}) = \sum_{l=0}^n \binom{n}{l} \underbrace{[\hat{p},[\cdots,[\hat{p},O(\hat{x})],\cdots],]}_{l \text{ commutators of } \hat{p}}   \hat{p}^{n-l}
  = \sum_{l=0}^n \binom{n}{l}  (-i)^l \partial_x^l O(\hat{x})  \hat{p}^{n-l}.
  \label{eq-operator-ordering}
\end{align}
Thus, any expectation value can be written in terms of expectation values of the ordered form $\langle \tilde{O}(\hat{x}) \hat{p}^n \rangle$, where $\tilde{O}(\hat{x})$ is a function of $\hat{x}$. Hence, we restrict our attention to relations among expectation values of this form. We will show that the expectation values $\langle O(\hat{x}) \hat{p}^n \rangle$ can be expressed in terms of expectation values of the form $\langle \tilde{O}(\hat{x}) \rangle$ by using the constraint equations \eqref{eq-HO=0} and \eqref{eq-HO=EO}. Therefore, the operator $\hat{p}$ can be eliminated from the set of independent variables.
\\

 To see this, we first use the relation \eqref{eq-HO=EO},
\begin{align} 
\langle OH \rangle = E \langle O \rangle.
\label{eq-HO=EO-appen}
 \end{align}
(This relation is equivalent to $\langle HO \rangle = E \langle O \rangle$ through $\langle [H,O] \rangle = 0$. However, we use the form \eqref{eq-HO=EO-appen} here because it is more convenient when dealing with ordered operators of the form $O(\hat{x}) \hat{p}^n$.) Applying the relation \eqref{eq-HO=EO-appen} with the Hamiltonian \eqref{eq-Hamiltonian-general} to the operator $O(\hat{x}) \hat{p}^n$, we obtain
\begin{align} 
\langle  O(\hat{x}) \hat{p}^n H \rangle = E \langle  O(\hat{x}) \hat{p}^n \rangle \quad \Longrightarrow \quad \langle O(\hat{x}) \hat{p}^{n+2} \rangle = 2E \langle O(\hat{x}) \hat{p}^n \rangle - 2 \langle O(\hat{x}) \hat{p}^n V(\hat{x}) \rangle .
\label{appen-eq:recursion-relations-HO-Op2}  
 \end{align}
Here, the last term $\langle O(\hat{x}) \hat{p}^n V(\hat{x}) \rangle$ can be expressed in terms of $\langle \tilde{O}(\hat{x}) \hat{p}^l \rangle$ $(0\le l \le n )$ by using the operator-ordering relation \eqref{eq-operator-ordering}. Therefore, $\langle O(\hat{x}) \hat{p}^{n+2} \rangle$ can be written as a linear combination of expectation values of the form $\langle \tilde{O}(\hat{x}) \hat{p}^l \rangle$ with $0\le l \le n $. Hence, by repeatedly applying this relation \eqref{appen-eq:recursion-relations-HO-Op2} starting from $n=0$, any expectation value $\langle O(\hat{x}) \hat{p}^n \rangle$ can be expressed solely in terms of $\langle \tilde{O}(\hat{x}) \rangle$ and $\langle \tilde{O}(\hat{x}) \hat{p} \rangle$.\\

Next, we consider the relation $\langle [H,O] \rangle = 0$ \eqref{eq-HO=0}. Applying this relation to the operator $O(\hat{x})$, we obtain
\begin{align} 
0=\langle [H, O(\hat{x})] \rangle = \frac{1}{2} \langle  [\hat{p}^2, O(\hat{x})] \rangle
\quad \Longrightarrow \quad & \langle [\hat{p}, O(\hat{x})] \hat{p} \rangle =- \frac{1}{2} \langle [\hat{p},[\hat{p},O(\hat{x}) ]]  \rangle . 
\label{const:OH-HO at k=0-com}\\
& \quad \left(  \braket{ \partial_x O (\hat{x}) \hat{p} }= \frac{i}{2} \braket{ \partial_x^2   O (\hat{x}) } \right)
\label{const:OH-HO at k=0}
 \end{align}
This relation allows us to express expectation values of the form $\langle \tilde{O}(\hat{x}) \hat{p} \rangle$ in terms of $\langle \tilde{O}(\hat{x}) \rangle$. By combining this with the recursion relation \eqref{appen-eq:recursion-relations-HO-Op2}, all expectation values involving $\hat{p}$ can therefore be written solely in terms of expectation values of the form $\langle \tilde{O}(\hat{x}) \rangle$.\\

So far, we have used the relation  $\langle OH \rangle = E \langle O \rangle$ \eqref{eq-HO=EO-appen} for operators of the form $O(\hat{x}) \hat{p}^n$ ($n \ge 0$), and the relation $\langle [H,O] \rangle = 0$ \eqref{eq-HO=0} for operators of the form $O(\hat{x})$. 
We now examine the role of the relation \eqref{eq-HO=0} when applied to operators of the form $O=O(\hat{x}) \hat{p}^{n}$ ($n \ge 1$). In fact, this relation yields new constraints on the expectation values $\langle \tilde{O}(\hat{x}) \rangle$ only for the case $n = 1$, whereas for $n \ge 2$ it does not provide additional independent constraints.

Let us first consider the case $n\ge 2$. Suppose that the relations $\langle [H,O(\hat{x}) \hat{p}^{l}] \rangle = 0$ for all $0\le l \le n-2$ are already known for arbitrary operators $O(\hat{x})$, and ask whether the relation $\langle [H,O(\hat{x}) \hat{p}^{n}] \rangle = 0$ provides any new constraint on the expectation values \cite{Aikawa:2021qbl}:
\begin{align} 
  0= \langle [H,O(\hat{x}) \hat{p}^{n}] \rangle =&
2\langle [H,O(\hat{x}) \hat{p}^{n-2} (H-V)  ] \rangle \nonumber \\
=&2 E \langle [H,O(\hat{x}) \hat{p}^{n-2}  ] \rangle  
-2  \langle [H,O(\hat{x}) \hat{p}^{n-2} V ] \rangle. 
\end{align}
Here we have used $\hat{p}^n=\hat{p}^{n-2} \hat{p}^2=2\hat{p}^{n-2}(H-V)$ and the relation \eqref{appen-eq:recursion-relations-HO-Op2}. In this expression, the last term $O(\hat{x}) \hat{p}^{n-2} V$ can be rewritten in terms of $\tilde{O}(\hat{x}) \hat{p}^{l}$ ($0 \le l \le n-2$) by using the operator-ordering relation \eqref{eq-operator-ordering}. Thus, every term on the right-hand side can be expressed as $\langle [H,\tilde{O}(\hat{x}) \hat{p}^{l}] \rangle$ ($0 \le l \le n-2$), all of which are already known by assumption. Therefore, the relation $\langle [H, O(\hat{x}) \hat{p}^{\,n}] \rangle = 0$ 
does not yield any new independent constraint.

Finally, we consider the case $n=1$. Applying the relation \eqref{eq-HO=0} to the operator $O(\hat{x})\hat{p}$, we obtain
\begin{align}
2 [\hat{p}, O(\hat{x})] \hat{p}^2 + [\hat{p},  [\hat{p}, O(\hat{x})]] \hat{p}  -  2[\hat{p}, V(\hat{x})] O(\hat{x}) =0.
\end{align}
In this equation, we eliminate the terms involving $\hat{p}^2$ and $\hat{p}$ by substituting the recursion relation \eqref{appen-eq:recursion-relations-HO-Op2} at $n=0$ and the relation \eqref{const:OH-HO at k=0-com}, where in both relations we replace $O(\hat{x})$ with $[\hat{p}, O(\hat{x})]$. This yields
\begin{equation}\label{master-eqn}
    \braket{\partial_x^3 O(\hat x)}+8 E\braket{ \partial_x O(\hat x)}-8\braket{(\partial_x O(\hat x))V(\hat x)}-4\braket{O(\hat{x})\partial_x V(\hat x)}=0.
\end{equation}
This provides relations among expectation values that do not involve the operator $\hat p$. As discussed above, all available constraints have now been taken into account. Consequently, to determine a minimal set of independent expectation values, it is sufficient to solve Eq.~\eqref{master-eqn}. The explicit form of this relation will be presented for the anharmonic oscillator and the mixed potential in the following subsections.

\subsection{Constraint equations in the anharmonic oscillator} 
\label{appen-sec:recursion-derivation-anharm}

We consider the anharmonic oscillator \eqref{eq-H-anharmonic} with the Hamiltonian
\begin{equation}
	H = \frac{1}{2}\hat{p}^2 + \frac{1}{2}\hat{x}^2 + \frac{1}{4}\hat{x}^4 \ .
\end{equation}
We focus on deriving the constraints for expectation values of operators of the form $\langle \hat{x}^m \hat{p}^n \rangle$, which constitute the building blocks of the bootstrap matrix in Eq.~\eqref{eq-operators-XP}.

First, from Eq.~\eqref{appen-eq:recursion-relations-HO-Op2} with $O=\hat{x}^m \hat{p}^n$, we obtain 
\begin{align}
\langle\hat x^m \hat p^{n+2} \rangle 
=&\;2 E \langle\hat x^m\hat p^n \rangle
- \langle\hat x^{m+2}\hat p^n \rangle
- \frac{1}{2}\langle\hat x^{m+4}\hat p^n \rangle+ 2i n \langle \hat x^{m+1}\hat p^{n-1} \rangle
\notag \\
&+2 i n \langle\hat x^{m+3}\hat p^{n-1} \rangle+ n(n-1)\langle\hat x^m\hat p^{n-2} \rangle
+ 3n(n-1)\langle\hat x^{m+2}\hat p^{n-2} \rangle \notag \\
&- 2i n(n-1)(n-2)\langle\hat x^{m+1}\hat p^{n-3} \rangle - \frac{1}{2} n(n-1)(n-2)(n-3)\langle\hat x^m\hat p^{n-4} \rangle,
\label{appen-eq:HO=EO-expand}
\end{align}
 where we have used the ordering relation \eqref{eq-operator-ordering},
\begin{align}
    \label{ordering-1}
	\hat{p}^k\hat{x}^m = \sum_{s=0}^{\text{min}(k,m)} \binom ks \frac{m!}{(m-s)!} (-i)^s \hat{x}^{m-s} \hat{p}^{k-s} \ .
\end{align}
From this relation, expectation values of the form $\langle \hat{x}^m \hat{p}^n \rangle$ with $n \geq 2$ can be recursively expressed in terms of the set $\{\langle \hat{x}^r \rangle, \langle \hat{x}^k \hat{p} \rangle\}$, where the values of $r$ and $k$ are determined by $m$ and $n$.\\

Next, the constraint \eqref{const:OH-HO at k=0-com} with $O=\hat{x}^{m+1}$ can be evaluated as
\begin{equation}
    \braket{\hat x^{m}\hat p}=\frac{i}{2}m\braket{\hat x^{m-1}}.
    \label{appen-eq:comm-HO-cosh-k=0-expand}
\end{equation}
This allows us to write all expectation values of the form $\braket{\hat x^m\hat p}$ in terms of $\braket{\hat x^{m-1}}$.

Finally, Eq.~\eqref{master-eqn} with $O=\hat{x}^m$ can be evaluated as
\begin{equation}
    \langle \hat{x}^{m+3} \rangle=\left(\frac{m}{2}+1\right)^{-1}\left(2mE \langle \hat{x}^{m-1} \rangle
+ \frac{1}{4}m(m-1)(m-2)\langle \hat{x}^{m-3}\rangle
- (m+1)\langle \hat{x}^{m+1} \rangle \right).
\label{appen-eq:recursion-anharmonic-xm}
\end{equation}
Since this expression is valid for all $m\ge0$, every expectation value $\braket{\hat x^l}$ with $l \ge 3$ can be expressed recursively in terms of the lower moments $\braket{\hat x}$ and $\braket{\hat x^2}$. Consequently, the independent variables for the anharmonic oscillator are 
\begin{equation}
    \{E,\langle \hat{x}\rangle,\langle \hat{x}^2\rangle\}.
    \label{eq-anharm-independent-variables}
\end{equation}
Here, we include the energy $E$ as an independent variable, since, in our bootstrap analysis, we scan over its possible values together with those of the other expectation values.

We may set $\langle \hat{x} \rangle = 0$ by applying the parity symmetry of the Hamiltonian. However, we do not impose this condition here, as we wish to maintain a consistent framework for comparison with the mixed potential.

\subsection{Constraint equations in the mixed potential} 
\label{appen-eq:recursion-derivation-mixed}

Let us now consider the case of the mixed potential, whose Hamiltonian is given by
\begin{equation}
	H = \frac{1}{2}\hat{p}^2 + (\hat{x}-L)^2 + \cosh \hat{x} = \frac{1}{2}\hat{p}^2 + (\hat{x}-L)^2 + \frac{1}{2}e^{\hat{x}} + \frac{1}{2}e^{-\hat{x}} \ .
\end{equation}
In this case, we construct the bootstrap matrix using the operator set \eqref{eq-operators-XEP}
\begin{equation}
	O =O(\hat x)\hat p^k= \hat{x}^m e^{n\hat{x}} \hat{p}^k,
\end{equation}
and we focus on deriving the corresponding constraint equations for these operators.  
We will show that the number of independent variables increases with the moments and, in fact, becomes unbounded.  
This behavior stands in sharp contrast to the anharmonic oscillator case.
\\

First, from Eq.~\eqref{appen-eq:recursion-relations-HO-Op2} with $O=\hat{x}^m e^{n\hat{x}}\hat{p}^{k}$, we obtain
\begin{align}
\langle \hat{x}^m e^{n\hat{x}}\hat{p}^{k+2} \rangle
=& \,2(E-L^2) \langle \hat{x}^m e^{n\hat{x}}\hat{p}^k \rangle  - 2\langle \hat{x}^{m+2} e^{n\hat{x}} \hat{p}^k\rangle +4L \langle \hat{x}^{m+1} e^{n\hat{x}}\hat{p}^k\rangle + 4 i k \langle \hat{x}^{m+1} e^{n\hat{x}}\hat{p}^{k-1}\rangle\notag \\& -4L i k \langle \hat{x}^m e^{n\hat{x}}\hat{p}^{k-1}\rangle  + 2k(k-1)\langle \hat{x}^m e^{n\hat{x}}\hat{p}^{k-2}\rangle- \sum_{r=0}^k \binom{k}{r}(-i)^r
\langle \hat{x}^m e^{(n+1)\hat{x}} \hat{p}^{k-r}\rangle\notag \\
&  - \sum_{r=0}^k \binom{k}{r}i^r
\langle \hat{x}^m e^{(n-1)\hat{x}} \hat{p}^{k-r}\rangle. \label{appen-eq:HO=EO-cosh-expanded}
\end{align}
Here we have used the ordering relation \eqref{eq-operator-ordering},
\begin{equation}
    \hat{p}^k e^{n\hat{x}} = \sum_{s=0}^{k} \binom ks (-in)^s e^{n\hat{x}} \hat{p}^{k-s},
\end{equation}
together with the relation \eqref{ordering-1}.
Using Eq.~\eqref{appen-eq:HO=EO-cosh-expanded}, any expectation value of the form $\braket{ \hat{x}^me^{n\hat{x}}\hat{p}^{k}}$ with $k\ge 2$ can be recursively expressed in terms of the set $ \{\braket{\hat x^q e^{r \hat x}},\braket{\hat x^s e^{t \hat x}\hat p}\}$, where the indices $q,r,s$ and $t$ are determined by $m,n$ and $k$.\\

Next, we consider the relation \eqref{const:OH-HO at k=0} for $O=\hat{x}^m e^{n\hat{x}}$, which can be computed as
\begin{equation}
	in \langle \hat{x}^{m}e^{n\hat{x}}\hat{p} \rangle = - im \langle \hat{x}^{m-1}e^{n\hat{x}}\hat{p} \rangle - mn \langle \hat{x}^{m-1}e^{n\hat{x}} \rangle- \frac{1}{2}m(m-1) \langle \hat{x}^{m-2}e^{n\hat{x}}\rangle - \frac{1}{2}n^2 \langle \hat{x}^{m}e^{n\hat{x}} \rangle \ . 
     \label{appen-eq:comm-HO-cosh-k=0}
\end{equation}
As in the anharmonic oscillator case, this relation allows us to express  
$\langle \hat{x}^{m} e^{n\hat{x}} \hat{p} \rangle$  
in terms of expectation values that do not involve $\hat{p}$.  
To achieve this, we must rewrite the first term $\langle \hat{x}^{m-1}e^{n\hat{x}}\hat{p} \rangle$ on the right-hand side. This is done by repeatedly applying the relation \eqref{appen-eq:comm-HO-cosh-k=0} with successively smaller values of $m$, starting from $m-1$ and continuing down to $m = 0$, for which we have
\begin{equation}
	\langle e^{n\hat{x}} \hat{p} \rangle = \frac{i}{2}n \langle e^{n\hat{x}}\rangle.  \label{appen-eq:comm-HO-cosh-k=0-m=0}
\end{equation}
Therefore, the expectation value $\langle \hat{x}^{m}e^{n\hat{x}}\hat{p} \rangle$ can be expressed in terms of the set $\{ \langle \hat{x}^l e^{n\hat{x}} \rangle \}$, ($0\leq l \leq m$), and the operator $\hat{p}$ has been fully eliminated.\\

Finally, we consider the constraint relations among the expectation values $\{ \langle \hat{x}^m e^{n\hat{x}} \rangle \}$. 
Substituting $O(\hat x)=\hat x^m e^{n\hat x}$ into Eq.~\eqref{master-eqn}, we obtain 
\begin{align}
8n \left\langle \hat{x}^{m+2} e^{n\hat{x}} \right\rangle = &
(16nL-8m-8)\left\langle \hat{x}^{m+1}e^{n\hat{x}}\right\rangle
+(n^3+8En-8nL^2+16mL+8L)\left\langle \hat{x}^{m}e^{n\hat{x}}\right\rangle
\notag\\
&+m(3n^2+8E-8L^2)\left\langle \hat{x}^{m-1}e^{n\hat{x}}\right\rangle+3m(m-1)n\left\langle \hat{x}^{m-2}e^{n\hat{x}}\right\rangle \notag\\
&+m(m-1)(m-2)\left\langle \hat{x}^{m-3}e^{n\hat{x}}\right\rangle \notag\\
& -4m\left\langle \hat{x}^{m-1}e^{(n+1)\hat{x}}\right\rangle
-(4n+2)\left\langle \hat{x}^{m}e^{(n+1)\hat{x}}\right\rangle \notag\\&-4m\left\langle \hat{x}^{m-1}e^{(n-1)\hat{x}}\right\rangle
-(4n-2)\left\langle \hat{x}^{m}e^{(n-1)\hat{x}}\right\rangle.
\label{eq-recursion-mixed}
\end{align}
We use this relation to systematically reduce the power of $\hat{x}$ appearing in the independent variables\footnote{This is one possible way to implement the constraint. Alternatively, one may use the relation to connect operators of the form $e^{\hat{x}}$ with different exponents, in which case the resulting set of independent variables would differ.}. For simplicity, we restrict ourselves to the case $n\neq0$. We observe that the power of $\hat{x}$ on the left-hand side is $m+2$, whereas on the right-hand side it ranges from $m-3$ to $m+1$. This structure allows us to solve the constraint recursively, starting from $m=0$, so that the expectation value $\langle \hat{x}^{m+2} e^{n\hat{x}} \rangle$ can be expressed entirely in terms of expectation values involving at most $\hat{x}^1$.

We now examine the contribution of the terms $\hat{x}^m e^{(n\pm1)\hat{x}}$ appearing on the right-hand side. As the recursion proceeds, the power of  $\hat{x} $ is reduced by two every two steps, while the exponent of $e^{\hat x}$ is simultaneously shifted by $\pm 1$ through these terms. Taking this behavior into account, we find that $\langle \hat{x}^{m+2} e^{n\hat{x}} \rangle$ can be expressed in terms of the set of variables
\begin{equation}\label{independent in mixed}
    \{E,\braket{ e^{k \hat x}},\braket{\hat x e^{l \hat x}}\}, \quad n-\lfloor (m+2)/2 \rfloor \le k,l \le n+\lfloor (m+2)/2 \rfloor,
\end{equation}
where $\lfloor z \rfloor$ denotes the floor function, i.e., the greatest integer less than or equal to $z$.

At this stage, all available constraint equations have been exhausted, and no further reduction of the number of variables in Eq.~\eqref{independent in mixed} is possible. Consequently, the number of independent variables depends explicitly on $m$ and $n$. In particular, as $m$ increases, the number of independent variables also increases, and there is no upper bound on their total number. The same behavior occurs in the case $n = 0$.  
This stands in sharp contrast to the anharmonic oscillator case, for which any expectation value $\langle \hat{x}^m \hat{p}^n \rangle$ can be reduced to the fixed set of three independent variables given in Eq.~\eqref{eq-anharm-independent-variables}.

\section{Numerical bootstrap method} 
\label{appen-sec:bootstrap-method-detail}

This appendix elaborates on the numerical implementation of the bootstrap method discussed in the main text. 

\subsection{Anharmonic oscillator}
\label{app-num-AHO}

We discuss the numerical implementation of the conventional bootstrap method for the anharmonic oscillator \eqref{eq-H-anharmonic}.
The elements of the conventional bootstrap matrix ${\mathcal M}$ in Eq.~\eqref{anharmonic-bootstrap-matrix} take the form
\begin{equation}\label{linear-moments}
    \langle \hat p^k\hat x^l\hat p^m\rangle=\sum_{0\leq n\leq 2} \mathcal{C}_n^{klm}(E)\langle \hat x^n\rangle,
\end{equation}
where the coefficients $\mathcal{C}_{n}^{klm}(E)$ are polynomials in the energy eigenvalue $E$.  
This expression is obtained by using the ordering relation \eqref{ordering-1} together with the recursion relations  
\eqref{appen-eq:HO=EO-expand}, \eqref{appen-eq:comm-HO-cosh-k=0-expand}, and \eqref{appen-eq:recursion-anharmonic-xm},  
as discussed in Appendix~\ref{appen-sec:recursion-derivation-anharm}.  
Although deriving the explicit form of $\mathcal{C}_n^{klm}(E)$ by hand is cumbersome, it can be computed straightforwardly using symbolic manipulation in \textit{Mathematica}.

In the bootstrap method, we determine the possible values of $E$ that are consistent with the positive semidefinite condition ${\mathcal M} \succeq 0$. The region of such values is referred to as the allowed region, from which the energy eigenvalues of the system can be extracted.

To identify the allowed region, we use the fact that the matrix elements of the bootstrap matrix \eqref{linear-moments} are linear in $\langle \hat x \rangle$ and $\langle \hat x^2 \rangle$, while the coefficients $\mathcal{C}_n^{klm}(E)$ are polynomial functions of $E$.
Thus, once a particular value of the energy, say $E = E_0$, is substituted into the bootstrap matrix, the problem of finding values of $\langle \hat x \rangle$ and $\langle \hat x^2 \rangle$ that satisfy ${\mathcal M} \succeq 0$ reduces to a semidefinite programming problem, which can be solved efficiently using numerical optimization techniques. 
If no values of $\langle \hat x \rangle$ and $\langle \hat x^2 \rangle$ satisfy the condition ${\mathcal M} \succeq 0$ for $E = E_0$, then $E_0$ is excluded from the allowed region.
By repeating this procedure over a range of $E$ values, we obtain the allowed region for the energy.

A sample implementation of this procedure is provided in Appendix~B of Ref.~\cite{Aikawa:2025dvt}, and the corresponding code is available at \cite{Morita:code}. \\

In our numerical analysis, we derive the bootstrap matrix elements \eqref{linear-moments} and construct ${\mathcal M}$ using \textit{Mathematica}  (the same procedure is applied to the mixed potential case). We then employ MOSEK through the \texttt{cvxpy} library in Python to determine the maximum and minimum values of $\langle \hat x \rangle$ for a given value of $E$ by solving the corresponding semidefinite programming problem. The tolerance of the solver is set to $10^{-10}$. The obtained allowed regions are shown in Fig.~\ref{fig-x4-x} and summarized in Table~\ref{table:energy-bounds-aho}. (We do not analyze bounds on $\langle \hat x^2 \rangle$, as the allowed region in the $(\langle \hat x \rangle, E)$ plane already provides sufficient information about the properties of the system.)

\paragraph*{Numerical errors} In our numerical analysis of the allowed regions at a fixed size of the bootstrap matrix ${\mathcal M}$, two parameters play a central role in controlling the precision of the results:  
the bin size of the energy grid and the tolerance of the solver.

A smaller bin size yields a more precise determination of the allowed region, but at the cost of increased computational time.  
In practice, it is sufficient to choose a bin size that is small compared with the characteristic width of the allowed region.

The tolerance of the solver determines how accurately the optimal solution to the semidefinite programming problem is obtained.  
A smaller tolerance generally leads to more accurate solutions.  
For moderate bootstrap matrix sizes, the dependence on the tolerance is mild.  
However, for larger matrices, the results can become sensitive to the tolerance, and numerical noise may appear.  
In such cases, the choice of solver can also influence the outcome.  
In this work, we use the CSDP and MOSEK solvers in \textit{Mathematica}, as well as MOSEK in Python.  
We compare the performance of different solvers and select the one that provides the most reliable results for each computation.
\\

\subsection{Mixed potential} 

We now consider the mixed potential \eqref{eq-H-coshx}.
The bootstrap matrix \eqref{eq-bootstrap-matrix} is constructed using the operator set \eqref{eq-operators-XEP}, and its matrix elements take the form
\begin{equation}\label{normal-ordered}
    \langle \hat p^k \hat x^{i}e^{j \hat x} \hat p^{l} \rangle=\sum_{ n }\mathcal{C}_n^{kijl} (E)\langle  e^{n \hat x}  \rangle+\sum_{ q }\tilde{\mathcal{C}}_q^{kijl}(E)\langle\hat x e^{q\hat x}\rangle,
\end{equation}
where the ranges of the summations over $n$ and $q$ depend on the indices $i,j,k$ and $l$ in a manner analogous to Eq.~\eqref{independent in mixed}. The coefficients $\mathcal{C}_{n}^{kijl}(E)$ and $\tilde{\mathcal{C}}_{q}^{kijl}(E)$ are polynomials in the energy $E$.

To obtain the allowed regions, we apply the same semidefinite-programming-based procedure as in the anharmonic oscillator case. The only difference is that, due to the structure of Eq.~\eqref{independent in mixed}, the number of independent variables is not fixed but instead depends on the size of the bootstrap matrix.

In our numerical analysis, we use \texttt{SemidefiniteOptimization} with the CSDP solver in \textit{Mathematica} to solve the resulting semidefinite programming problem. The tolerance of the solver is set to $10^{-6}$. The allowed regions obtained in this manner are shown in Figs.~\ref{fig-cosh-bootstrap}, \ref{fig-cosh-bootstrap-shifted}, and \ref{fig-cosh-L=0-bootstrap} (left panel).

However, the results near the lower bounds of $E$ obtained with the CSDP solver exhibit numerical noise, and we therefore use the MOSEK solver in \textit{Mathematica} to compute the values reported in Table~\ref{table:energy-bounds-mixed}. (We do not use MOSEK for the results shown in Figs.~\ref{fig-cosh-bootstrap}, \ref{fig-cosh-bootstrap-shifted} and \ref{fig-cosh-L=0-bootstrap}, because the MOSEK solver becomes noisy for larger values of $E$. In practice, MOSEK performs better than CSDP only for smaller values of $E$. In this way, the choice of solver is made case by case to obtain the most reliable results.)

\subsubsection{Parity constraint} \label{appen-sec:parity}

As discussed in Section~\ref{sec-parity-condition}, we can use the parity condition Eq.~\eqref{eq-parity-condition} to resolve the underdetermination in the $L=0$ case. 
We impose this condition on the matrix elements \eqref{normal-ordered} and perform the semidefinite programming analysis.

We use the same numerical method (the CSDP solver in Mathematica) as in the case without imposing the parity condition.
The obtained allowed regions are shown in Fig.~\ref{fig-cosh-L=0-bootstrap}.

\subsubsection{Additional positivity condition} \label{appen-sec:add-pos-condition}

As discussed in Section~\ref{sec-additional-inequalities}, we can use the positive semidefinite condition ${\mathcal M}^{\cosh} \succeq 0$ to resolve the underdetermination in the bootstrap analysis. We determine the allowed region that simultaneously satisfies both ${\mathcal M} \succeq 0$ and ${\mathcal M}^{\cosh} \succeq 0$ by applying the numerical semidefinite programming method.

In the numerical analysis, we use the same method (MOSEK in Python) as in the anharmonic oscillator case.
The tolerance of the solver is set to $10^{-6}$. The obtained allowed regions are shown in Figs.~\ref{fig-cosh-L=-1-bootstrap} and \ref{fig:cosh-2mat-L=0-bootstrap}, and the corresponding bounds are summarized in Tables~\ref{table:energy-bounds-add-matrix} and \ref{table:energy-bounds-add-matrix-parity}.

\section{Resolving the Sign Underdetermination in $e^{\hat{x}}$} \label{appen-sign}

In this appendix, we discuss resolutions of the underdetermination of the $\pm$ sign in the operator $e^{\hat{x}}$ in the conventional bootstrap method.
Note that the prescriptions discussed in Section \ref{sec-taylor-expansion} and Section \ref{sec-additional-inequalities} may also solve the sign underdetermination.
Here, we propose two other possible methods.
We do not implement these methods numerically in this paper because the prescriptions introduced in Section~\ref{sec-ambiguity-resolution} are already sufficient to reproduce the spectrum of the mixed potential. Nevertheless, the methods proposed here may be useful in future studies.

The first method is to take $P=e^{\hat{x}}$ in Eq.~\eqref{eq-positivity} and construct a new bootstrap matrix 
\begin{align}
	{\mathcal M}_{ ij}^{\exp}:=
		\left\langle O_i^\dagger e^{\hat{x}} O_j \right\rangle .
\end{align}
Then, by imposing the positive semidefinite condition ${\mathcal M}^{\exp} \succeq 0$ in addition to the conventional bootstrap condition ${\mathcal M} \succeq 0$, we explicitly impose the positivity of the operator $e^{\hat{x}}$ and thereby resolve its sign underdetermination.

The second method is to use the operator set
\begin{align} 
 &   \{\hat{x}^l  e^{m\hat{x}/2} \hat{p}^n  \}, \quad l=0,1, \cdots, K_x,~m=-K_{\rm e}, \cdots, K_{\rm e},~ n=0,1, \cdots, K_p,  &(K_e \ge 2).  
 \label{eq-operators-XEP-2}
 \end{align}
instead of the one used in Eq.~\eqref{eq-operators-XEP} to construct the bootstrap matrix. 
Then, although $e^{\hat{x}/2}$ itself may be replaced by $\pm e^{(\hat{x}-x_0)/2}$, its square is \[ \left(\pm e^{(\hat{x}-x_0)/2}\right)^2 =e^{\hat{x}-x_0}, \] so that the sign freedom disappears for $e^{\hat{x}}$.\\

We emphasize that, although these two methods remove the sign freedom, the parameter $x_0$ remains undetermined.

{\normalsize 
\bibliographystyle{utphys}
\bibliography{QM} }

\end{document}